# Mid-infrared Colors Vary with Galactic Environment: Contrasting Star-forming Disks, Young Centers, and Quiescent Star-formation Deserts

Debosmita Pathak[1,2,3], Karin M. Sandstrom[4], Adam K. Leroy[1,2], Ryan Chown[5], Daniel A. Dale[6], Hannah B. Koziol[4], Mansi Padave[4], Jessica Sutter[7], Thomas G. Williams[8], Dalya Baron[9,10], Alberto D. Bolatto[11], Médéric Boquien[12], Oleg V. Egorov[13], Eric Emsellem[14,15], Simon C. O. Glover[16], Ralf S. Klessen[16,17], Eva Schinnerer[18], Todd A. Thompson[1,2,19], Francesco Belfiore[20], Frank Bigiel[21], Enrico Congiu[22], Ivan Gerasimov[12], Jay González Lobos[18], Annie Hughes[23], Laura A. Lopez[1,2], Justus Neumann[18], Elias K. Oakes[24], Hsi-An Pan[25], Erik Rosolowsky[26], and Sumit K. Sarbadhicary[27]

[1] Department of Astronomy, Ohio State University, 180 W. 18th Ave., Columbus, OH 43210, USA; pathak.89@buckeyemail.osu.edu
[2] Center for Cosmology and Astroparticle Physics, 191 West Woodruff Ave., Columbus, OH 43210, USA
[3] IPAC, California Institute of Technology, 1200 E. California Blvd., Pasadena, CA 91125, USA
[4] Department of Astronomy & Astrophysics, University of California, San Diego, 9500 Gilman Dr., La Jolla, CA 92093, USA
[5] Faculty of Computer Science & Technology, Algoma University, Sault Ste. Marie, ON P6A 2G4, Canada
[6] Department of Physics and Astronomy, University of Wyoming, Laramie, WY 82071, USA
[7] Whitman College, 345 Boyer Ave., Walla Walla, WA 99362, USA
[8] UK ALMA Regional Centre Node, Jodrell Bank Centre for Astrophysics, Department of Physics and Astronomy, The University of Manchester, Oxford Rd., Manchester M13 9PL, UK
[9] Kavli Institute for Particle Astrophysics & Cosmology, Stanford University, CA 94305, USA
[10] Center for Decoding the Universe, Stanford University, CA 94305, USA
[11] Department of Astronomy and Joint Space-Science Institute, University of Maryland, College Park, MD 20742, USA
[12] Université Côte d'Azur, Observatoire de la Côte d'Azur, CNRS, Laboratoire Lagrange, 06000, Nice, France
[13] Astronomisches Rechen-Institut, Zentrum für Astronomie der Universität Heidelberg, Mönchhofstraße 12-14, D-69120 Heidelberg, Germany
[14] European Southern Observatory, Karl-Schwarzschild-Straße 2, 85748, Garching, Germany
[15] Université Lyon 1, ENS de Lyon, CNRS, Centre de Recherche Astrophysique de Lyon UMR5574, 69230, Saint-Genis-Laval, France
[16] Universität Heidelberg, Zentrum für Astronomie, Institut für Theoretische Astrophysik, Albert-Ueberle-Str. 2, 69120, Heidelberg, Germany
[17] Universität Heidelberg, Interdisziplinäres Zentrum für Wissenschaftliches Rechnen, Im Neuenheimer Feld 225, 69120 Heidelberg, Germany
[18] Max-Planck-Institut für Astronomie, Königstuhl 17, D-69117, Heidelberg, Germany
[19] Department of Physics, Ohio State University, 91 West Woodruff Ave., Columbus, OH 43210, USA
[20] INAF—Osservatorio Astrofisico di Arcetri, Largo E. Fermi 5, I-50157, Firenze, Italy
[21] Argelander-Institut für Astronomie, Universität Bonn, Auf dem Hügel 71, 53121 Bonn, Germany
[22] European Southern Observatory (ESO), Alonso de Córdova 3107, Casilla 19, Santiago 19001, Chile
[23] IRAP/OMP/Université de Toulouse, 9 Av. du Colonel Roche, BP 44346, F-31028 Toulouse cedex 4, France
[24] Department of Physics, University of Connecticut, 196A Auditorium Rd., Storrs, CT 06269, USA
[25] Department of Physics, Tamkang University, No.151, Yingzhuan Rd., Tamsui District, New Taipei City 251301, Taiwan
[26] Department of Physics, 4-183 CCIS, University of Alberta, Edmonton, AB T6G 2E1, Canada
[27] Department of Physics and Astronomy, The Johns Hopkins University, Baltimore, MD 21218, USA
*Received 2026 April 21; revised 2026 June 26; accepted 2026 July 13; published 2026 August 24*

## Abstract

We present 50–100 pc-resolution JWST/MIRI and NIRCam measurements of mid-infrared (mid-IR) color variations in the diffuse interstellar medium (ISM) of 71 nearby star-forming galaxies from the PHANGS-JWST survey. Mid-IR emission traces the dust column density, intensity ($U$) and hardness of the interstellar radiation field, and the physical state (charge, size) and abundance of polycyclic aromatic hydrocarbons (PAHs). Mid-IR colors that trace PAH band ratios remain fairly constant in the diffuse ISM of star-forming disks. However, they show stark variations in extreme environments: highly star-forming central molecular zones (CMZs) and star formation deserts/quiescent bulges. In CMZs, PAH-to-continuum (3.3/21, 7.7/21, and 11.3/21 $\mu$m) and the 10/21 $\mu$m continuum colors are 0.2–0.4 dex lower than in normal disks. We attribute this to higher $U$ based on the far-infrared dust colors and the high 21 $\mu$m/$\Sigma_{\rm Mol}$, which we suggest to be a good tracer of $U$ outside star-forming regions. Meanwhile, star formation deserts show low 7.7 $\mu$m PAH emission, resulting in low 7.7/21 $\mu$m and 7.7/11.3 $\mu$m, while all other mid-IR colors remain typical. This suggests the presence of more neutral PAHs in star formation deserts, where low 7.7 $\mu$m likely reflects ISM conditions similar to early-type and elliptical galaxies. All environments form part of a continuous trend in 7.7/11.3 $\mu$m versus specific star formation rate.

## 1. Introduction

The mid-infrared (mid-IR) captures emission from polycyclic aromatic hydrocarbons (PAHs)—small carbonaceous nanoparticles with prominent emission features across the near- and mid-IR (A. Leger & J. L. Puget 1984). PAHs are ubiquitous in the interstellar medium (ISM) of the Milky Way and nearby galaxies, and typically account for ∼10%–20% of the total IR luminosity from galaxies (J. D. T. Smith et al. 2007; A. G. G. M. Tielens 2008). PAHs play a crucial role in ISM physics, especially in regulating photoelectric heating. Low PAH abundance or the presence of highly charged PAHs results in less efficient photoelectric heating, which can impact the ISM phase structure (M. G. Wolfire et al. 1995, 2003). PAH emission encodes information on the abundance of PAHs, total dust column density, local radiation field (intensity and hardness), as well as the physical state (charge, size) of PAHs (L. J. Allamandola et al. 1989; E. Peeters et al. 2002; B. van Diedenhoven et al. 2004; A. G. G. M. Tielens 2008). PAH emission has thus also been used as a tracer of star formation (e.g., E. Peeters et al. 2004; A. G. G. M. Tielens 2005; B. Gregg et al. 2026), radiation field (E. L. O. Bakes & A. G. G. M. Tielens 1994; R. Genzel et al. 1998; D. Rigopoulou et al. 1999; D. Baron et al. 2025), and recently cold gas (A. K. Leroy et al. 2023; D. Pathak et al. 2024; R. Chown et al. 2025).

Several prominent PAH emission features are accessible to JWST, e.g., at 3.3, 6.2, 7.7, 8.6, 11.3, 12.7, and 17 $\mu$m, which trace stretching and bending modes from C − C to C − H in PAH molecules (e.g., L. J. Allamandola et al. 1999). Variations in the relative strengths of these PAH features can also reflect differences in PAH size distribution (smaller PAHs emit strongly in the shorter-wavelength features compared to larger PAHs emitting strongly at longer wavelengths), charge (e.g., 7.7 $\mu$m from ionized PAHs versus 3.3 and 11.3 $\mu$m from more neutral PAHs) and physical conditions of the ISM (e.g., radiation field intensity and hardness; E. L. O. Bakes & A. G. G. M. Tielens 1994; B. T. Draine & A. Li 2001; P. Panuzzo et al. 2011; A. Maragkoudakis et al. 2020; B. T. Draine et al. 2021; D. Baron et al. 2025; D. A. Dale et al. 2025; B. Gregg et al. 2026).

PAH emission is sensitive to changes in the interstellar radiation field (ISRF) shape and intensity, especially the availability of energetic far-ultraviolet (FUV) photons that efficiently heat PAHs. Observations show that PAH survival and properties vary in response to their environment, as evidenced by relative variations among their strong spectral features. PAH emission depends on galactic environment (J. D. T. Smith et al. 2007; S. Stierwalt et al. 2013; A. Li 2020), PAHs are destroyed in H II regions (K. M. Sandstrom et al. 2010; J. Sutter et al. 2024; O. V. Egorov et al. 2025), and can be suppressed at lower metallicity (C. W. Engelbracht et al. 2005; B. O'Halloran et al. 2006; B. T. Draine et al. 2007; F. Galliano et al. 2008; L. K. Hunt et al. 2010; C. M. Whitcomb et al. 2024; R. Chown et al. 2025), while their formation pathway is currently being investigated (e.g., A. M. Burkhardt et al. 2021; C. Xue et al. 2025).

Mid-IR colors and PAH properties can vary on <10 pc scales in bright H II regions and on $\lesssim$100 pc scales in galaxy centers, where $U$, the ISM radiation field intensity, can be orders of magnitude higher than the $U \sim 1$ across most parts of galaxies (e.g., G. Helou et al. 2004; D. A. Dale et al. 2005; B. T. Draine et al. 2007; J. C. Muñoz-Mateos et al. 2009). Constraining variations in PAH properties in the diffuse ISM thus requires (1) high-resolution data that can separate diffuse emission from <10 pc-scale ionizing sources, and (2) wide-area coverage that spans a range of galactic environments. Characterizing mid-IR colors and PAH property variations in the diffuse ISM of large samples of nearby galaxies is now possible with JWST. JWST's subarcsecond resolution allows us to separate near- and mid-IR emission from nebular regions with very high $U$ from the surrounding diffuse medium across galaxies out to at least ∼20 Mpc (e.g., J. Chastenet et al. 2023; A. K. Leroy et al. 2023; K. M. Sandstrom et al. 2023; D. Pathak et al. 2024; J. Sutter et al. 2024; R. Chown et al. 2025; O. V. Egorov et al. 2025 studying subsets of our sample). The field of view of its MIRI and NIRCam imaging instruments allows it to cover large parts of a nearby galaxy in a single observation.

Building on previous work within the Physics at High Angular resolution in Nearby GalaxieS (PHANGS[28]) Collaboration (J. Chastenet et al. 2023; D. Baron et al. 2024, 2025; J. Sutter et al. 2024; R. Chown et al. 2025; D. A. Dale et al. 2025; O. V. Egorov et al. 2025; H. Kolziol et al. 2026,in preparation; ; see Section 4.1), we provide the first systematic JWST-MIRI measurements of diffuse ISM mid-IR colors as a function of large-scale environment (centers, bars, bulges, disks) in 71 star-forming galaxies mapped by JWST, and critically examine the role of environment in driving diffuse dust emission. We use mid-IR band ratios to investigate variations in PAH properties, abundance, and heating in the diffuse ISM. By studying a large set of galaxies, we aim to (1) measure the typical PAH-related mid-IR colors in the diffuse ISM of star-forming galaxies, (2) identify environments where the diffuse ISM deviates from these typical conditions, and (3) diagnose the drivers of these changes by comparison to other bands and properties of the local environment. Our results are representative of the diffuse ISM in "normal" star-forming galaxies that span the star-forming main sequence, where interpreting changes in mid-IR colors and PAH emission systematically takes into account the local environment.

The paper is organized as follows. In Section 2, we describe our sample, JWST and MUSE data, data processing, and environment classification. In Section 3, we present our inventory of mid-IR radial profiles and colors in normal disks and galaxy centers, including central molecular zones (CMZs) and bulges. In Section 4, we interpret the observed mid-IR color variations due to galactic environment. Finally, we summarize our key conclusions in Section 5.

## 2. Data and Methods

Our sample consists of 71 nearby star-forming galaxies from the PHANGS (Physics at High Angular resolution in Nearby GalaxieS) surveys (A. K. Leroy et al. 2021; J. C. Lee et al. 2023; R. Chown et al. 2025) that have high-resolution (10–50 pc-scale) multi-wavelength data. The full list of galaxies and associated measurements have been included as a machine-readable table, as described in Table 3. This includes JWST NIRCam and MIRI imaging for all 71 galaxies, and MUSE optical IFU coverage for 42 galaxies (E. Emsellem et al. 2022; O. V. Egorov et al. 2025). The 71 galaxies span the star-forming main sequence with stellar masses between $10^{9.16}$ and $10^{11.14}\,M_\odot$ and global star formation rate (SFR) between 0.15 and 16.90 $M_\odot\,\mathrm{yr}^{-1}$. The sample captures a wide range of local environments representative of normal star-forming galaxies. Global properties of the full sample are summarized in A. K. Leroy et al. (2021).

[28] https://phangs.stsci.edu/

### 2.1. PHANGS-JWST

We use near- and mid-IR images from the PHANGS-JWST Cycle 1 (GO 2107, PI: Lee; J. C. Lee et al. 2023) and Cycle 2 Treasury (GO 3707, PI: Leroy; R. Chown et al. 2025, A. Leroy et al. 2026, in preparation). We focus on the diffuse ISM using five filters: NIRCam F300M (3 $\mu$m) primarily traces starlight; MIRI F770W (7.7 $\mu$m) traces C–C stretching modes of more ionized PAHs; F1000W (10 $\mu$m) traces a combination of continuum from stochastically heated small grains, silicate absorption, and PAH emission; F1130W (11.3 $\mu$m) traces C–H out-of-plane bending modes of more neutral PAHs; and F2100W (21 $\mu$m) traces the hot dust (small grains) continuum (A. K. Leroy et al. 2023; C. M. Whitcomb et al. 2023; G. P. Donnelly et al. 2024; R. Chown et al. 2025c). In Section 4.4 we also compare to maps of 3.3 $\mu$m PAH emission captured in the F335M filter for the 19 Cycle 1 galaxies. This feature traces smaller, more neutral PAHs. H. Koziol et al. (submitted) describe the production of these $F335M_{PAH}$ feature maps.

The Cycle 1 PHANGS Treasury includes 19 galaxies with full coverage of all filters. We include 52 galaxies from the Cycle 2 Treasury with F300M, F770W (the shorter-wavelength PAH band), and F2100W.[29] NGC 4826 (Cycle 2) has additional F1000W and F1130W coverage from GO 3177 (PI Sun), which brings our full sample to 20 galaxies with full MIRI coverage, and 51 galaxies without F1000W and F1130W.

The data are reduced using `pjpipe` (T. G. Williams et al. 2024). Modifications to `pjpipe` used in the latest combined Cycle 1 and 2 data are described in R. Chown et al. (2025).[30] In brief, `pjpipe` adds modifications to the STScI JWST pipeline (H. Bushouse et al. 2025) to improve the handling of extended sources. The JWST observations are astrometrically registered against Hubble Space Telescope broadband imaging using bright point sources, which in turn have been aligned to Gaia DR3 sources.

We analyze the NIRCam and MIRI images after convolving them to a common $0\farcs9$ Gaussian PSF, which is effectively beam-matched to the MUSE data resolution. For F2100W, our longest wavelength filter, this resolution also represents a compromise between improving signal-to-noise (S/N) and decreasing angular resolution (see T. G. Williams et al. 2024; R. Chown et al. 2025 for details).

#### 2.1.1. Background Determination

We study the systematic variations in ratios between mid-IR filter intensities in the diffuse ISM. This makes determining the background level important, but most of our JWST images include little or no empty sky. Following A. K. Leroy et al. (2023), the backgrounds at F300M and F770W are "anchored" to match wide-field mapping at similar wavelengths (NIRCam F300M to WISE1, MIRI F770W to WISE3) and then the backgrounds in JWST images at other bands are anchored to match those images. Since all MIRI filters (F1000W, F1130W, F2100W) are anchored to F770W, which sets the intercept between F770W and each MIRI filter to zero, zero-level offsets are not a concern for our measurements. The resulting MIRI backgrounds are uncertain by $\pm$0.03–0.1 MJy sr$^{-1}$ ($\sim$0.03 MJy sr$^{-1}$ in F770W, $\sim$0.09 MJy sr$^{-1}$ in F2100W; see T. G. Williams et al. 2024). Since the NIRCam F335M maps suffer from known striping issues,[31] we restrict all comparisons with $F335M_{PAH}$ (only in Section 4.4) to high-confidence regions where $F335M_{PAH}$ is greater than the stripe intensity ($\sim$0.04 MJy sr$^{-1}$; see H. Koziol et al. submitted).

#### 2.1.2. Starlight Subtraction

The MIRI filters usually contain some stellar continuum emission in addition to PAH and/or small dust continuum emission (see, e.g., G. Helou et al. 2004; D. A. Dale et al. 2009; C. M. Whitcomb et al. 2023; J. Sutter et al. 2024). We estimate and remove contamination from the stellar continuum for each MIRI filter using the NIRCam F300M emission as a template for the stellar emission. We use the prescription from J. Sutter et al. (2024) for F300M to scale and subtract stellar emission from F770W. For the longer-wavelength MIRI filters, we follow J. Sutter et al. (2024) and use spectral energy distribution models produced by the Code for Investigating Galaxy Emission (`CIGALE`; M. Boquien et al. 2019) to determine the factor to scale F300M to predict starlight emission in the filter. The formulae that we use are

$$\begin{aligned}
\mathrm{F770W_{ss}} &= \mathrm{F770W} - (0.22 \pm 0.08) \times \mathrm{F300M},\\
\mathrm{F1000W_{ss}} &= \mathrm{F1000W} - (0.14 \pm 0.02) \times \mathrm{F300M},\\
\mathrm{F1130W_{ss}} &= \mathrm{F1130W} - (0.11 \pm 0.02) \times \mathrm{F300M},\\
\mathrm{F2100W_{ss}} &= \mathrm{F2100W} - (0.032 \pm 0.005) \times \mathrm{F300M}.
\end{aligned}$$

We use the intensity in each filter in MJy sr$^{-1}$, the "ss" subscript indicates the starlight-subtracted intensity. We include the $\pm 1\sigma$ uncertainty in starlight subtraction from the standard deviation from the `CIGALE` models computed over a variety of plausible `CIGALE` models (see J. Sutter et al. 2024) in each MIRI filter. Starlight subtraction makes the largest difference in regions with low MIRI intensities and high F300M, such as near-IR bright bulges and star formation deserts (see Section 2.3). Our sample includes 13 galaxies with such old stellar bulges and bars with very high surface densities of old stellar populations. In these regions, we find that stellar continuum accounts for $\lesssim$30% of the flux in F770W and as much as $\sim$5% of the flux in F2100W. In such regions, the starlight subtraction can represent the dominant source of uncertainty.

#### 2.1.3. Estimating $R^*_{\mathrm{PAH}}$

With starlight-subtracted mid-IR maps, we measure the mid-IR color of $\mathrm{F770W_{ss}}$ to $\mathrm{F2100W_{ss}}$ across the full sample. $\mathrm{F770W_{ss}/F2100W_{ss}}$ is often employed to trace the PAH fraction in normal star-forming galaxies (e.g., J. Sutter et al. 2024; O. V. Egorov et al. 2025), following similar use of the 8/24 $\mu$m ratio with Spitzer (e.g., C. W. Engelbracht et al. 2005; F. Galliano et al. 2018; A. Li 2020). We estimate the mid-IR PAH-to-continuum emission ratio $R^*_{\mathrm{PAH}}$ using the empirical calibration[32] from J. Sutter et al. (2024) as

$$R^*_{\mathrm{PAH}} = 2.57\frac{\mathrm{F770W_{ss}}}{\mathrm{F2100W_{ss}}}. \quad (1)$$

[29] Three galaxies from the Cycle 2 Treasury are excluded due to issues with establishing a robust background level for the images.

[30] For more details on updates to `pjpipe`, please refer to https://pjpipe.readthedocs.io/en/latest/.

[31] https://jwst-docs.stsci.edu/known-issues/1-f-noise

[32] The diffuse $R^*_{\mathrm{PAH}}$ measured using Equation (1) is consistent with the slope of the best-fit line to the $\mathrm{F770W_{ss}} - \mathrm{F2100W_{ss}}$ relation.

Note that while J. Sutter et al. (2024) and O. V. Egorov et al. (2025) did not subtract starlight from F2100W, we use $F2100W_{ss}$ in $R^{*}_{PAH}$. As noted above, this results in a discernible $\lesssim$5% effect only in extremely near-IR bright old bulges and star formation deserts.

Note that while we remove the stellar continuum, we do not subtract dust continuum emission from PAH-dominated filters $F770W_{ss}$ and $F1130W_{ss}$ since adjacent mid-IR filters are not available to reliably estimate the dust continuum in PAH filters. In previous work, the hot dust continuum is estimated to be $\sim$5× lower than the 7.7 $\mu$m PAH feature contribution to $F770W_{ss}$ (C. M. Whitcomb et al. 2023; D. Baron et al. 2024). Hence, we calculate other band ratios including $F770W_{ss}/F1130W_{ss}$ (tracing PAH charge) without dust continuum subtraction. For consistency, we also compare our results to dust models in which we include the continuum emission. Future work will present improved dust continuum constraints for MIRI filters for PHANGS galaxies (L. Hands et al. 2026, in preparation).

In Figure 1, we show the diversity in resulting maps of F300M, $F770W_{ss}$, $F2100W_{ss}$, and $R^{*}_{PAH}$ for a few representative galaxies. All intensity maps share the same colorbar, and similarly all $R^{*}_{PAH}$ panels share the same colorbar for visual comparison between filters and across galaxies.

### 2.2. PHANGS-MUSE and Archival MUSE Data

Previous works have found mid-IR colors to vary with radiation field intensity and hardness (e.g., J. Sutter et al. 2024; D. Baron et al. 2025; O. V. Egorov et al. 2025). We use maps of SFR and specific star formation rate (sSFR$\equiv$SFR/$M_*$) as tracers of these factors. SFR and sSFR reflect the presence and relative abundance of young massive stars, which produce the majority of FUV photons in most star-forming galaxies and so drive changes in radiation field intensity and spectrum. We measure the sSFR using optical integral field unit (IFU) spectroscopy from the Multi Unit Spectroscopic Explorer (MUSE) on the Very Large Telescope (VLT) available for 42/71 of our galaxies. This includes the 19 Cycle 1 galaxies with MUSE data from the PHANGS-MUSE Large Program (1100.B-0651, PI: Schinnerer; E. Emsellem et al. 2022). The PHANGS-MUSE data coverage for the 19 galaxies extends out to $\sim 2 \times R_{\rm eff}$ or twice the effective radii, which is generally larger than the PHANGS-JWST footprint. MUSE data on an additional 23 galaxies are available from a combination of several small programs and archival data at ESO (see Table 1 in O. V. Egorov et al. 2025, for details of specific archival observations). This constitutes the "MUSE Extended" sample, and brings our total sample with existing MUSE data up to 42. Within the MUSE Extended sample, roughly half of the galaxies have coverage comparable to PHANGS-MUSE (full optical disk), while coverage for the other half is restricted to single central pointings (usually $\sim 1 R_{\rm eff}$).

We use the "convolved and optimized" (copt) resolution MUSE data, which achieve a uniform Gaussian point-spread function (PSF) for each galaxy (as in E. Emsellem et al. 2022). The angular resolution of the MUSE cube varies by galaxy and is $\sim$0.″8–1″ for our sample. We treat these data as effectively beam-matched to the 0.″9 JWST working resolution, since we only use them for SFR and sSFR measurements across azimuthal bins or morphological environments, i.e., regions much larger than a resolution element.

As in E. Emsellem et al. (2022), both the PHANGS-MUSE and MUSE Extended sample are reduced using the PHANGS-MUSE pipeline `pymusepipe`[33] to produce data cubes, and the PHANGS data analysis pipeline `DAP`[34] to produce maps of SFR and stellar mass (see E. Emsellem et al. 2022, for details). `DAP` uses the MUSE Balmer decrement extinction-corrected H$\alpha$ maps to trace the star formation surface density, $\Sigma_{\rm SFR}$ (E. Emsellem et al. 2022; F. Belfiore et al. 2023a), and models the stellar continuum (e.g., I. Pessa et al. 2022, 2023) to trace the stellar mass surface density, $\Sigma_*$, which are reliable on $\gtrsim$80 pc scales, and certainly on environment-integrated scales, but can break down on smaller scales in $\lesssim$50 pc-scale H II regions or high-attenuation centers (E. Emsellem et al. 2022). The ratio of the two is sSFR $= \Sigma_{\rm SFR}/\Sigma_*$. When measuring sSFR in large regions, we sum total SFR and stellar mass in the region before dividing.

For the 32 galaxies in our sample without MUSE coverage, we use F2100W and F2100W/F300M as proxies for $\Sigma_{\rm SFR}$ and sSFR. We expect this approach to be less accurate than the MUSE products, but it has the advantage of being available for all targets. We discuss this more in Section 2.4.

### 2.3. Environmental Classification

We measure the variation in mid-IR colors as a function of morphological environment. We use the M. Querejeta et al. (2021) environmental masks that define galaxy centers, bars, and disk regions based on Spitzer IRAC near-IR imaging (see also H. Salo et al. 2015). Within each environment (center/bar/disk), we calculate the integrated $\Sigma_{\rm SFR}$ and sSFR as described above. Based on this, we identify four different types of environments[35]:

1. *Old stellar population-dominated star formation deserts or bulges.* Galaxy centers or bar environments (inter-bar regions) with low sSFR (logF2100W/F300M $\leqslant -0.3$, or sSFR $\lesssim 7 \times 10^{-12}$ yr$^{-1}$), hereby referred to as "bulges."
2. *Young star-forming centers.* Galaxy centers and CMZs[36] with high $\Sigma_{\rm SFR}$ (median F2100W intensity$>$10 MJy sr$^{-1}$, or $\Sigma_{\rm SFR} \gtrsim 0.03\, M_\odot$ yr$^{-1}$ kpc$^{-2}$).
3. *Normal centers.* Galaxy centers with $\Sigma_{\rm SFR}$ and sSFR typical of the rest of the galaxy.
4. *Disks.* All environments not classified as a galaxy center or "bulge," with $\Sigma_{\rm SFR}$ and sSFR typical of the rest of the galaxy. This includes, e.g., weak bars, spiral arms, and inter-arm regions.

While we use the M. Querejeta et al. (2021) morphological masks to distinguish between galaxy centers, bulges, and disks as our fiducial environment selection, we note three special cases. In NGC 4826, our MIRI and NIRCam observations cover the central $\approx 2\ell_*$, where $\ell_*$ is the exponential scale length of the stellar disk (H. Salo et al. 2015; A. K. Leroy et al. 2021).[37] NGC 4826 is close to an early-type galaxy (ETG; SAab Hubble type), where the inner

[33] https://github.com/emsellem/pymusepipe

[34] https://gitlab.com/francbelf/ifu-pipeline

[35] We check for mid-IR color variations between arms and inter-arms and do not see a noticeable difference, and hence include both within a common "disk" environment.

[36] In our sample, high SFR galaxy centers are rich in molecular gas (see D. R. Gleis et al. 2026).

[37] We find the $\ell_*$ derived from WISE1 and Spitzer 3.6 $\mu$m exponential fits from A. K. Leroy et al. (2021) are in good agreement with $M_*$ radial profiles derived from NIRCam F300M and a variable M/L ratio (as in A. K. Leroy et al. 2019; D. Pathak et al. 2025) for our sample of galaxies.

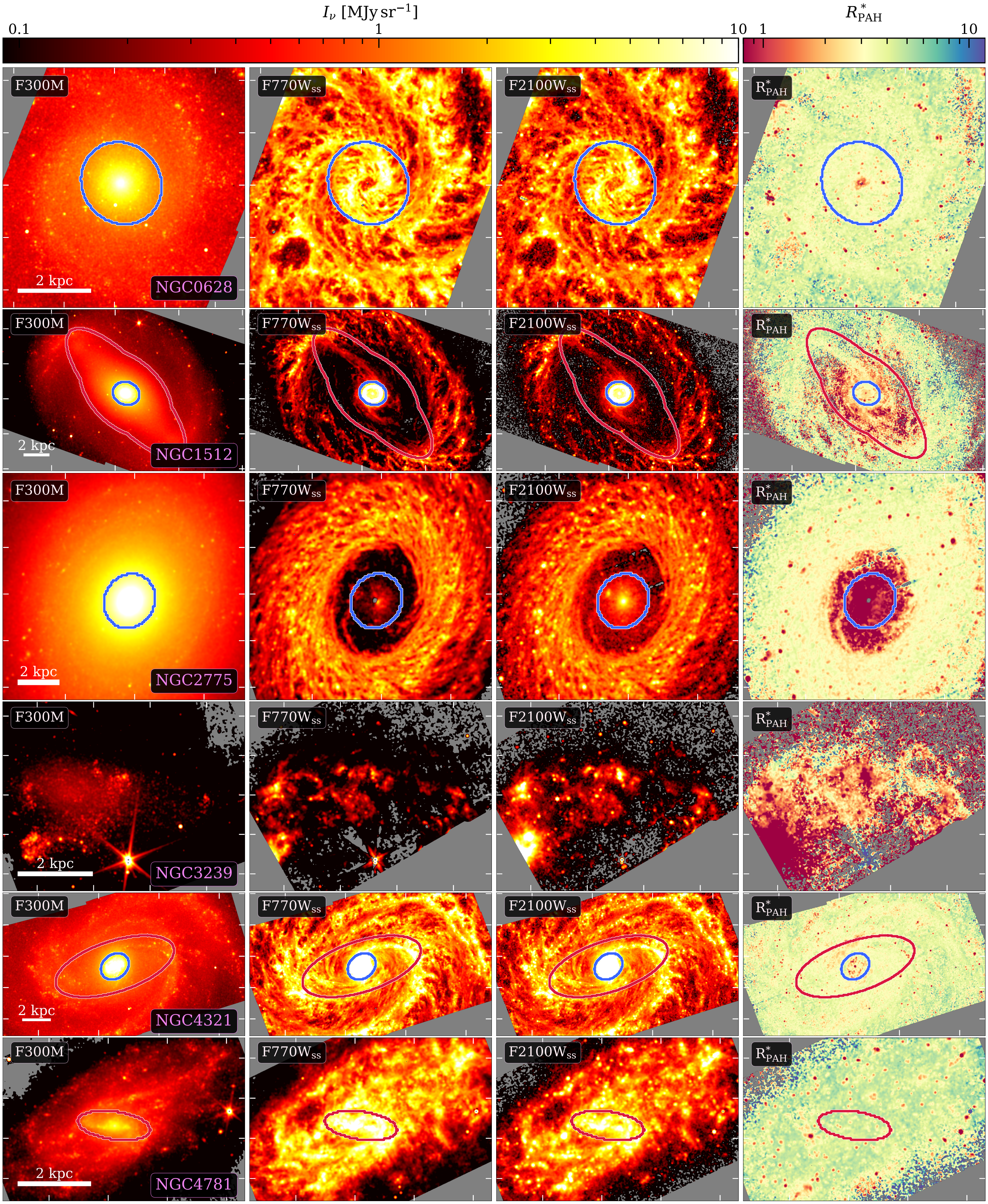

**Figure 1.** Zoom-ins showing a few example galaxies from our sample in F300M, F770W$_{\rm ss}$, F2100W$_{\rm ss}$, and $R^*_{\rm PAH}$, with M. Querejeta et al. (2021) morphological environment masks for centers (blue) and bars (red) overlaid for reference.

$\ell_*$ is star-forming with high gas column density within an early-type disk galaxy where a "recent" merger likely caused the counter-rotating H I disk dominated by older stellar populations at $R_{\rm gal} \gtrsim \ell_*$. Hence, in NGC 4826, in addition to the M. Querejeta et al. (2021) environment masks, we classify the outer $>\ell_*$ of the galaxy as a star formation desert. Finally, NGC 3368 and NGC 3344 were not included in M. Querejeta et al. (2021). Lacking environment masks, we visually classify the central $\ell_*$ as the "galaxy center" ($\ell_*$ measurement from A. K. Leroy et al. 2021), and $r > \ell_*$ as the "disk" environment for each galaxy.

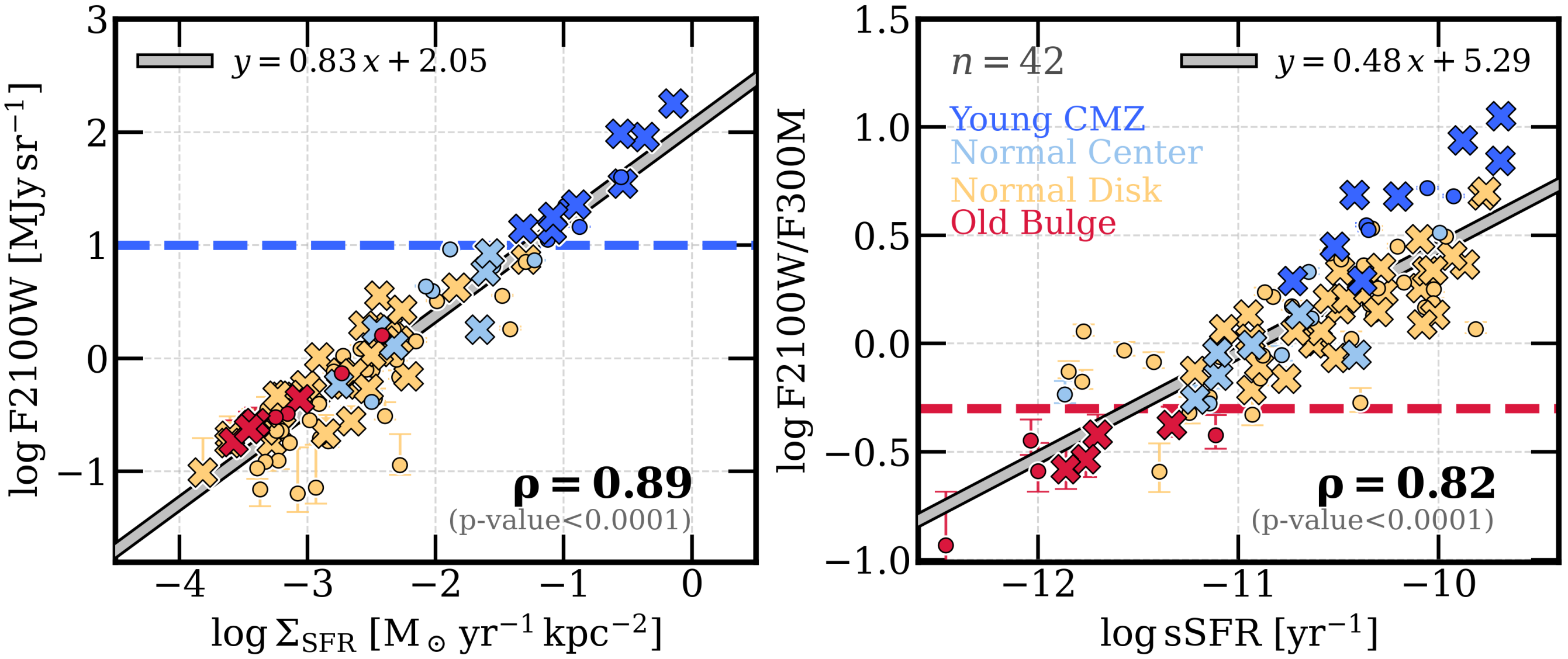

**Figure 2.** Correlating environment-integrated median $\Sigma_{\rm SFR}$ ($M_\odot$ yr$^{-1}$ kpc$^{-1}$; left panel) and sSFR (yr$^{-1}$; right panel) measurements from MUSE with JWST F2100W intensities and F2100W/F300M, respectively, for the 42 galaxies with joint MUSE and JWST coverage. Environments (Section 2.3) are indicated with colors—old bulges (red), young star-forming centers (blue), normal centers (light blue), and normal disks (light yellow) for targets with full MIRI filter coverage (20/71; X's) and partial filter coverage (51/71; circles). We include log−log fits to all environments (light gray); the corresponding correlation coefficient $\rho$ and p-value; and the threshold in F2100W (left) and F2100W/F300M (right) used to classify young centers (dark blue) and old bulges (red) as in Section 2.3 are indicated as horizontal dashed lines.

### 2.4. Using JWST Filters as $\Sigma_{\rm SFR}$ and sSFR Indicators

For the 32 galaxies without MUSE coverage, we use the median F2100W intensity in each environment as a $\Sigma_{\rm SFR}$ indicator. We also use the ratio of median F2100W-to-median F300M as a proxy for sSFR. MIRI F2100W is analogous to widely used IR SFR tracers including WISE4 and Spitzer MIPS 24 $\mu$m (e.g., R. C. Kennicutt & N. J. Evans 2012; T. H. Jarrett et al. 2013; A. K. Leroy et al. 2019) and has been explored as an SFR tracer using JWST including by F. Belfiore et al. (2023b); D. Calzetti et al. (2025). NIRCam F300M is analogous to the commonly used $M_*$ tracers WISE1 or IRAC1 (e.g., S. E. Meidt et al. 2014; M. Querejeta et al. 2015; A. K. Leroy et al. 2019) or the widely used $K$ band (e.g., T. H. Jarrett et al. 2003; M. F. Skrutskie et al. 2006).

In Figure 2, we show how F2100W and F2100W/F300M correlate strongly with $\Sigma_{\rm SFR}$ and sSFR for the 42 galaxies with MUSE coverage. The plot indicates strong correlations of $\Sigma_{\rm SFR}$ with F2100W (Spearman rank correlation coefficient $\rho = 0.89$) and sSFR with F2100W/F300M ($\rho = 0.82$) for the 42 galaxies with MUSE and JWST measurements, with $< 0.5$ dex scatter.[38] This supports our use of these IR intensities and colors as proxies for star formation activity where MUSE data are not available.

Using these tracers, we select young star-forming centers as galaxy center regions within the M. Querejeta et al. (2021) mask where the median F2100W intensity is $>$10 MJy sr$^{-1}$. This translates to roughly $\Sigma_{\rm SFR} > 0.03\,M_\odot$ yr$^{-1}$ kpc$^{-2}$. We select old starlight-dominated bulges and star formation deserts as galaxy center and bar regions with log F2100W/F300M $\leqslant -0.3$, which translates roughly to sSFR$\lesssim 7 \times 10^{-12}$ yr$^{-1}$. Both cuts are indicated in Figure 2 as dashed horizontal lines.

### 2.5. Core Measurements

#### 2.5.1. Constructing Radial Profiles

We measure the radial variation in dust emission, mid-IR colors, and $R^*_{\rm PAH}$ in the diffuse ISM in each galaxy. To consistently compare galaxies of different sizes and morphologies, we construct radial profiles in azimuthal bins of width $0.1\ell_*$, the exponential disk scale length. For reference, in a simple exponential disk, we expect that the effective radius relates to the scale length as $R_{\rm eff} = 1.68\ell_*$. However, the radial profiles of many of our galaxies cannot be fit by a single exponential due to the presence of CMZs or bars. We thus prefer $\ell_*$ instead of the $R_{\rm eff}$ or a half-light radius measured in the near-IR, since many of our galaxies show strong nuclear emission from active galactic nucleus (AGN) disks or nuclear stellar disks. In these cases, $R_{\rm eff}$ is affected by this bright compact emission, which results in smaller $R_{\rm eff}$ values, and is not a good predictor of the large-scale disk structure.

To construct radial profiles, we calculate the median intensities in each azimuthal $0.1\ell_*$ wide bin (Section 2.3). We use the median[39] to suppress the contribution of H II regions and properly access the diffuse ISM values. H II regions are bright, but occupy only a small fraction of the total area in any radial bin or environment mask (see Section 2.5.3 for exceptions), and typically represent only the tail end of the distribution of pixel intensities (D. Pathak et al. 2024). We demonstrate in Figure 3 that the ratio of medians is relatively insensitive to whether nebular regions are masked. We include the radial profiles of NGC 1300 since it is particularly illustrative—NGC 1300 includes all three major environments (young center/star formation desert/disk; Section 2.3), which are also discernible from the radial profile.

[38] Using either F2100W or F2100W$_{\rm ss}$ yields similar correlations.

[39] We note that the mode is consistent with the median (within 5%) due to the high angular resolution of MIRI and the large number of independent resolution elements thus included in each azimuthal bin.

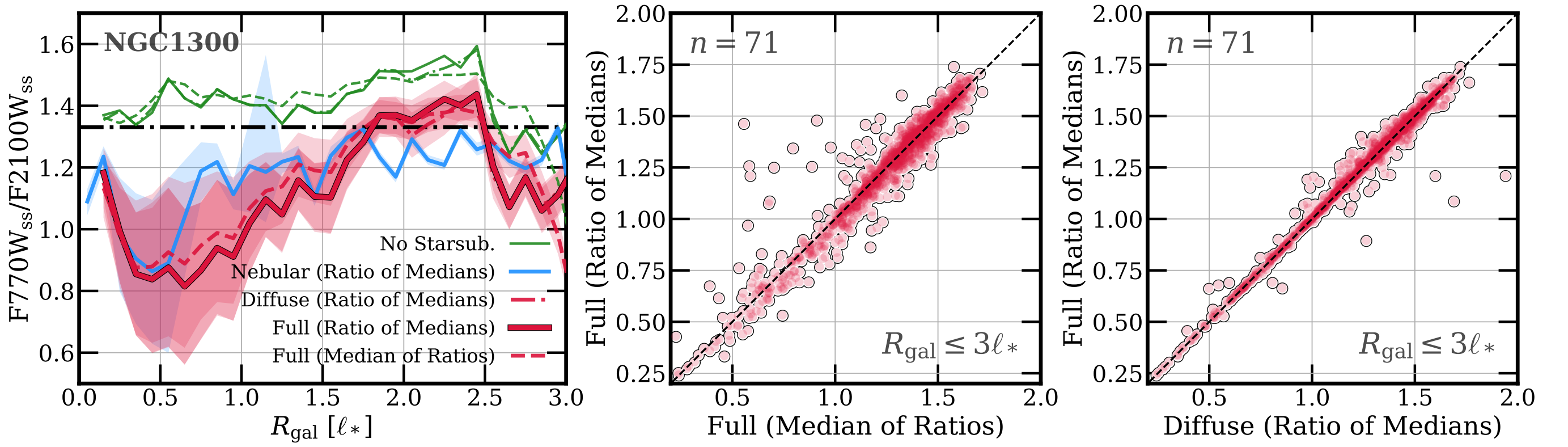

**Figure 3.** Left: constructing radial profiles of F770W$_{ss}$/F2100W$_{ss}$ for NGC 1300 as an example. The solid red line shows our fiducial measurement constructed from the ratio of medians taken across the full dataset. The other lines show the impact of varying the methodology. Green shows the profile without starlight subtraction for all pixels (solid green), for only diffuse emission (dotted–dashed green; see Section 2.5.2), and constructed taking the median of pixel-wise ratios instead of the ratio of medians (dashed green). Other red lines show varying methodologies with star-subtraction applied: considering diffuse emission only (dotted–dashed red), all pixels (solid red), or the median of pixel ratios (dashed red). Blue shows the ratio of median values for only nebular regions. The black dotted–dashed line shows the typical value for the diffuse emission from J. Sutter et al. (2024). Middle: testing agreement between the ratio of medians vs. the median of ratios for all radial bins, including both diffuse and nebular pixels, across 71 galaxies with $R_{gal} \leqslant 3\ell_*$. Right: testing agreement between the ratio of medians when including all pixels vs. including diffuse pixels only (masking nebular pixels before binning) for all radial bins at $R_{gal} \leqslant 3\ell_*$.

We measure the statistical uncertainty around the median using the error extension for JWST images (T. G. Williams et al. 2024) appropriately adjusted for the number of independent 0$''$.9 resolution elements in the bin. This statistical uncertainty is typically <5%. The backgrounds are uncertain to ∼0.03–0.09 MJy sr$^{-1}$, depending on filter (see Section 2.5.4), and this uncertainty is covariant across the entire profile.

We clip the profiles where the median intensity falls below 5× the propagated 1$\sigma$ error in the median in that bin, or falls to within 10% of the background intensity in a filter (e.g., where F2100W falls below $1.1 \times 0.09 \approx 0.1$ MJy sr$^{-1}$). This typically rejects radial bins in the outskirts where the emission is faint and the background is uncertain, such as corners of the mosaic. We also clip any bin at large $R_{gal}$ that contains <50 independent 0$''$.9 resolution elements, which again rejects corners of images where not enough azimuthal extent is covered in a bin. Finally, for galaxies with inclination >45° we do not include emission within ±30° of the minor axis when binning by radius, which is sufficiently conservative since the binned medians and percentiles are robust to <10% for ±10° variation in the azimuthal cut along the minor axis. This avoids line-of-sight confusion between the vertical and radial distributions of dust.

### 2.5.2. Measuring Mid-IR Colors of Diffuse Emission

Within azimuthal bins and kiloparsec-scale environments, we measure mid-IR colors. The ratio of medians is more resistant to outliers than the median of pixel-wise ratios, particularly in the faint, diffuse parts of disks. Therefore, we report all colors (intensity ratios) as

$$x/y = \frac{\text{median } x \text{ in environment}}{\text{median } y \text{ in environment}}. \quad (2)$$

In Figure 3 (middle), we show that the radial profiles of the ratios of medians are generally in agreement with the medians of pixel-wise ratios. We confirm this holds true for most radial bins at $R_{gal} \leqslant 3\ell_*$.

We also check whether our median-based ratios effectively pick out diffuse emission. Our primary concern is that our measurements might be affected by emission from small but bright H II regions where PAHs can be destroyed. To test this, we compare radially averaged bins of environment-integrated colors with and without masking nebular regions. We use the MUSE nebular region masks where available (B. Groves et al. 2023; O. V. Egorov et al. 2025), and otherwise denote all pixels with F2100W intensities >3 MJy sr$^{-1}$ as nebular, since the locations of H II regions typically show high F2100W intensities (D. Pathak et al. 2024).

In Figure 3 (right), we show that there is indeed good consistency between mid-IR colors with or without masking nebular regions. Since the nebular regions are only available for a subset of galaxies with MUSE data, and the MUSE-identified nebular masks are often larger than the actual physical size of the nebulae and include significant diffuse emission in addition to the central ∼10 pc-scale H II regions (see, e.g., A. T. Barnes et al. 2026), we prefer to treat the median as our fiducial measurement. This allows us to provide measurements that include all pixels in an environment while still reflecting diffuse emission in the ISM.

### 2.5.3. F2100W in Galaxy Centers

Young star-forming centers often host CMZs or AGN, which can saturate or contaminate the emission with strong diffraction spikes in F2100W. We visually inspect each galaxy center and remove from analysis the saturation- or diffraction artifact-dominated centers in NGC 1566, NGC 1637, NGC 3507, NGC 3627, NGC 4457, NGC 4597, NGC 4569, NGC 4941, NGC 5643, NGC 6300, and NGC 7496. We include an example of a center removed from the analysis in Figure 4.

Star-forming centers also host CMZs and nuclear starbursts, where, due to the high spatial density of nebular regions, the median no longer reliably excludes nebular regions. Even for galaxies with MUSE nebular catalogs, the MUSE data lack the resolution to distinguish H II regions from the surrounding medium. Instead, in the star-forming centers we separate nebular and diffuse emission using the JWST data. Correlating pixel-wise F2100W$_{\rm ss}$ with F770W$_{\rm ss}$ and $R^*_{\rm PAH}$ within each center, we fit a broken power law to binned medians (similarly to R. Chown et al. 2025). Each broken power-law fit yields $x_b$, the value of $x$, which corresponds to the break in the power law. Whenever F2100W$_{\rm ss} > x_b$, we consider the pixel nebular, where low F770W$_{\rm ss}$ and $R^*_{\rm PAH}$ reflect PAH destruction in ionized gas, and we consider all pixels $<x_b$ as diffuse. In Figure 5, we include an example galaxy center with the resulting selection for diffuse versus nebular pixels in NGC 1512.

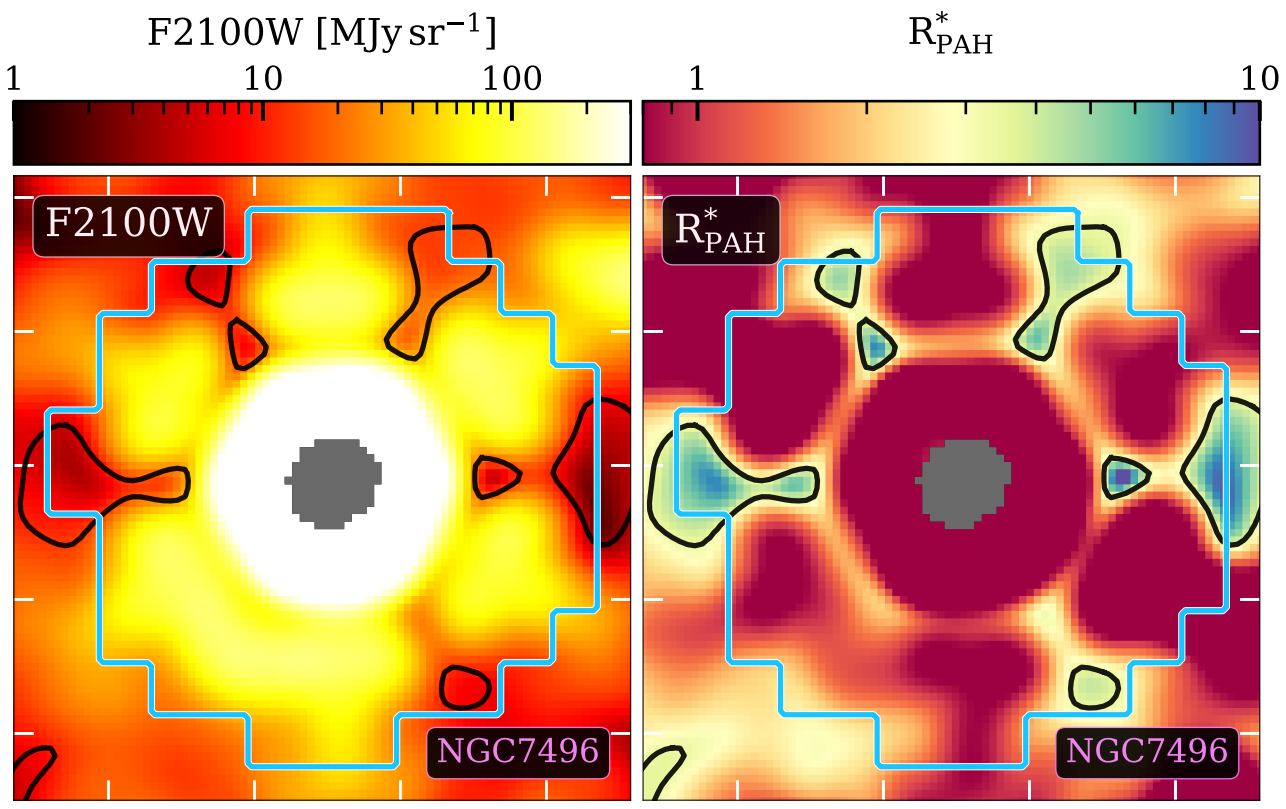

**Figure 4.** The center of NGC 7496, an example of galaxy centers removed from analysis due to contamination by prominent diffraction spikes in F2100W. We show maps of contaminated F2100W and resulting $R^*_{\rm PAH}$, blue contours for the center mask from M. Querejeta et al. (2021), and black contours for median $R^*_{\rm PAH}$ in the center.

In our analysis, we show two measurements for CMZs wherever possible—filled symbols for all pixels in centers (diffuse + nebular), and empty triangles for diffuse-only emission identified using this broken power-law approach. We take the diffuse measurement as our fiducial value for comparisons with other environments.

#### *2.5.4. Uncertainties*

Most emission in the inner parts of galaxies is detected at high S/N ($>5\sigma$), and each azimuthal ring or kiloparsec-scale environment contains a large number ($\gg$50) of 0$''$.9 independent resolution elements, which means the statistical error on the median is small ($\lesssim$1%). The innermost annuli, where the number of resolution elements is smaller, tend to be the brightest regions and the emission here is also detected at the highest S/N (typically $\gg 10\sigma$).

In this case, where statistical errors are insignificant, the uncertainty in our measurements is driven by systematics. Calibration uncertainties for JWST photometry are $\sim$5% (J. Rigby et al. 2023). Adding the uncertainty for two filters in quadrature, this translates to a $\sim$7% instrumental uncertainty on colors measured from JWST.

The uncertainty in determining MIRI backgrounds also affects our measurements in faint parts of galaxies and at large Galactocentric radii, where the median intensity can be of order the background. Some of our galaxies include empty sky regions, where we estimate the statistical noise at our common resolution for each MIRI filter. At 0$''$.9 resolution, typical $1\sigma$ uncertainties in the backgrounds are roughly 0.03, 0.04, 0.05, and 0.09 MJy sr$^{-1}$ at F770W, F1000W, F1130W, and F2100W, respectively. Thus, anchoring for MIRI images is uncertain to $\lesssim$0.04 MJy sr$^{-1}$ in F770W, F1000W, and F1130W, and $\lesssim$0.1 MJy sr$^{-1}$ in our longest wavelength filter F2100W.

For regions of galaxies at intermediate brightness ($I_\nu \approx$ 1MJy sr$^{-1}$), the uncertainty associated with stellar continuum subtraction is often the dominant source of uncertainty. From Section 2.1.2, the $1\sigma$ uncertainty is $\sim$15%–35% on the factor to scale the F300M and predict the starlight intensity in the MIRI filters. This uncertainty is largest in regions dominated by older stellar populations, i.e., bulges and star formation deserts, and at shorter wavelengths, i.e., F770W and F1000W. Our final uncertainties in all figures in Section 3 and beyond thus add in quadrature this dominant uncertainty due to starlight subtraction to the filter-dependent background uncertainty and the 5% calibration uncertainty for filter intensities, or 7% calibration uncertainty in mid-IR colors.

Finally, we provide the Spearman rank correlation coefficient $\rho$ between measurements where relevant, and print the corresponding $p$-value only if $>$0.0001. We print $\rho$ in bold if the $p$-value $<$ 0.0001, or the correlation is highly significant. For all correlations, we perform $2 \times 10^4$ permutations (or randomization tests) to estimate robust nonparametric $p$-values for small samples ($n \approx$ 40–150) which do not require assumptions of normality.

## 3. Mid-IR Colors in Galaxies

We use maps of 3 $\mu$m starlight, starlight-subtracted 7.7 $\mu$m, and 21 $\mu$m dust continuum emission for all 71 galaxies, and include 10 $\mu$m and 11.3 $\mu$m where available (20/71). We discuss three key results below: (1) Radial profiles of galaxies in the mid-IR show systematic variations with galaxy $M_*^{\rm total}$; (2) Mid-IR colors and $R^*_{\rm PAH}$ (scaled F770W$_{\rm ss}$/F2100W$_{\rm ss}$; Section 2.1.3) remain relatively constant in “normal” star-forming parts of galaxies; and (3) Mid-IR colors show stark environmental differences between normal star-forming disks, star-forming CMZs, and the quiescent bulges of massive galaxies.

### *3.1. Radial Profiles of 71 Galaxies*

In Figure 6, we compile the radial profiles of median F300M, F770W$_{\rm ss}$, and F2100W$_{\rm ss}$ for all 71 galaxies, and F1000W$_{\rm ss}$ and F1130W$_{\rm ss}$ for 20 galaxies following Section 2.5.1. We first show the profiles of each galaxy colored by $M_\star$ and also show median profiles after binning the galaxies by global $M_*$. More massive galaxies ($M_* > 10^{10}\,M_\odot$) show the imprint of central bulges and bars in their steeper or flattened central radial profiles (departures from an exponential disk) in addition to higher overall intensities, compared to lower-mass galaxies which show flatter profiles (E. Emsellem et al. 2026).

In Figure 7, the profiles of $R^*_{\rm PAH}$ show systematic variation with galaxy $M_*$. The amount and scale of $R^*_{\rm PAH}$ suppression in centers varies by galaxy $M_*^{\rm total}$, where the most massive galaxies ($M_* > 10^{10}\,M_\odot$) on median show the lowest $R^*_{\rm PAH}$ in centers. The central $R^*_{\rm PAH}$ in these massive galaxies is on median 0.1–0.2 dex lower than the nearly flat median beyond $\ell_*$ (discussed further in Section 3.3).

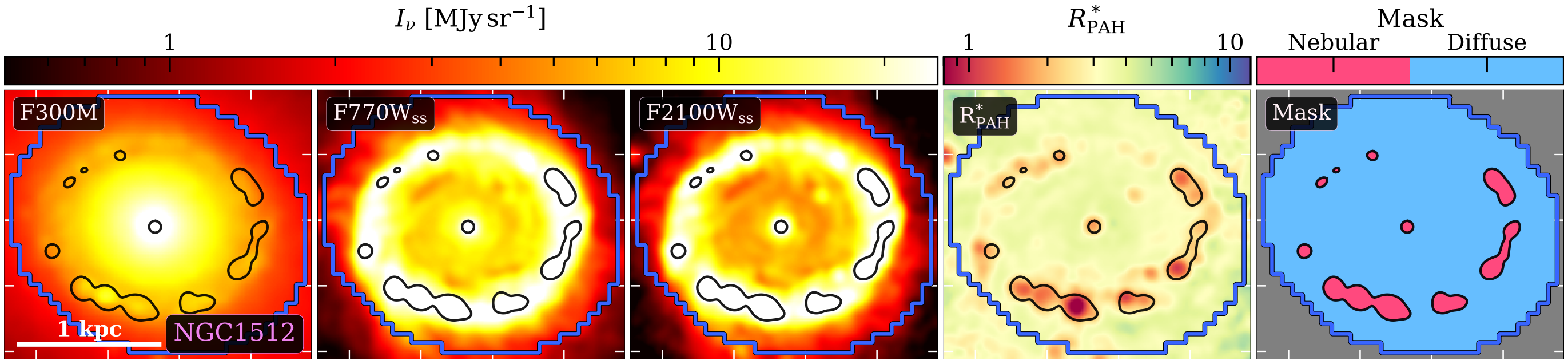

**Figure 5.** From left to right, F300M, F770W$_{\rm ss}$, F2100W$_{\rm ss}$, $R^*_{\rm PAH}$, and resulting masks for diffuse (blue) vs. nebular emission (pink) in the center of NGC 1512, with dark blue contours for the morphological environment mask for the center of NGC 1512 from M. Querejeta et al. (2021).

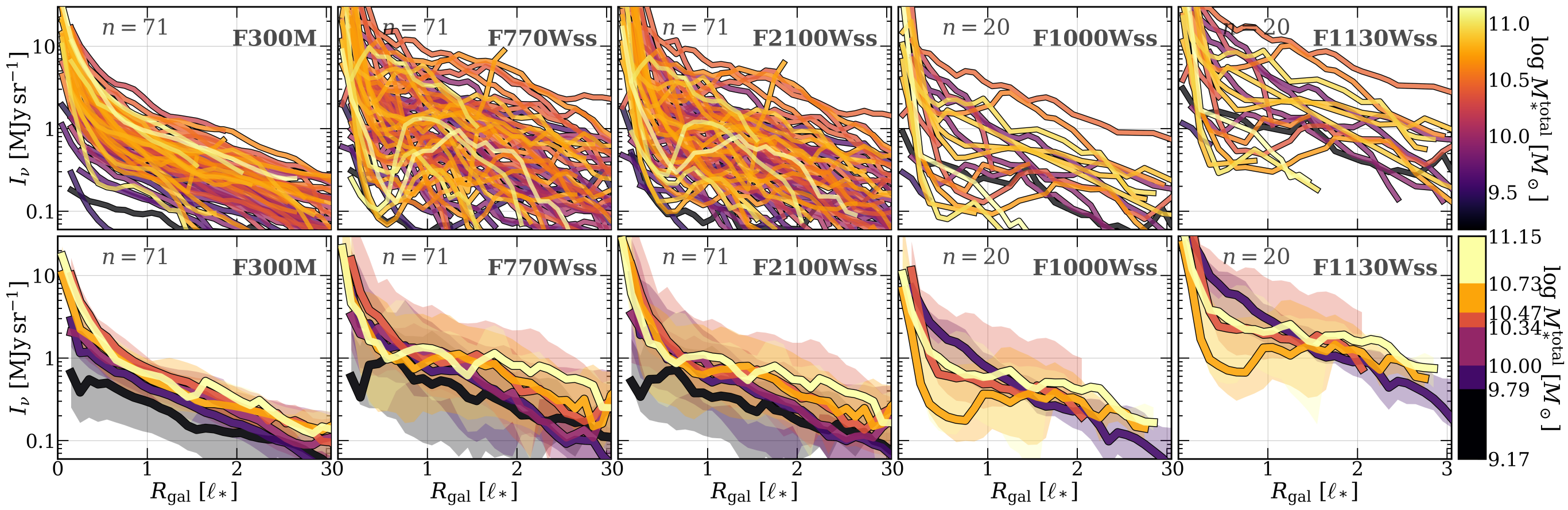

**Figure 6. Top**: radial profiles of F300M, F770W$_{\rm ss}$, and F2100W$_{\rm ss}$ intensity for all 71 galaxies; F1000W$_{\rm ss}$ and F1130W$_{\rm ss}$ for 20 galaxies, colored by galaxy stellar mass $M_*$. Bottom: median (solid lines) and 16th–84th percentile range (shaded region) of radial profiles of F300M, F770W$_{\rm ss}$, F2100W$_{\rm ss}$, F1000W$_{\rm ss}$, and F1130W$_{\rm ss}$ in six percentile bins of $M_*$. All radial profiles are shown as a function of (projected) Galactocentric radius in units of the galaxy exponential scale length $\ell_*$.

$R^*_{\rm PAH}$ remains nearly constant across most radial bins in the star-forming disks of "normal" galaxies beyond $\sim 1\ell_*$. This is reflected in both the individual and stacked radial profiles of $R^*_{\rm PAH}$, which remain approximately flat on average at $\ell_* < R_{\rm gal} < 3\ell_*$ for most galaxy disks (discussed further in Section 3.2).

Our NIRCam footprints typically extend out to only $\sim$2.5–$3\ell_*$, and rarely cover significant area beyond $4\ell_*$. Figure 7 shows that the profiles of $R^*_{\rm PAH}$ and $R^*_{\rm PAH}$ remain flat over this range, but our observations do not cover the outskirts of star-forming disks, where $R^*_{\rm PAH}$ is expected to vary due to changes in metallicity, sSFR, and lower column density (see, e.g., G. Aniano et al. 2020; C. M. Whitcomb et al. 2024; R. Chown et al. 2025). We thus do not cover enough radial extent and range in galaxy $M_*$ and metallicity to recover any systematic trend in $R^*_{\rm PAH}$ due to changing metallicity in the diffuse ISM.

### 3.2. Roughly Constant $R^*_{\rm PAH}$ in Normal Star-forming Disks

Table 1 summarizes our measurements of $\log_{10} R^*_{\rm PAH}$, which is defined in Equation (1). The radial profiles of $R^*_{\rm PAH}$ for most galaxies show a stable plateau between roughly 1 and $3\ell_*$. In Figure 8, we show this plateau value as the median over all $0.1\ell_*$ wide radial bins with $\ell_* \leqslant R_{\rm gal} \leqslant 3\ell_*$, as well as the $1\sigma$ scatter across azimuthal bins. We check against global (galaxy-integrated) properties (from A. K. Leroy et al. 2021) for correlations between the plateau value of $R^*_{\rm PAH}$. We find that the plateau value correlates best with global sSFR = SFR$^{\rm total}$/M$_*^{\rm total}$, with a moderate Spearman rank correlation coefficient of $\rho = 0.42$ ($p$-value$< 0.0001$). While galaxy sSFR presents the strongest correlation, we also note that the lowest $M_*^{\rm total}$ galaxies show the lowest $R^*_{\rm PAH}$ (no overall correlation with $M_*^{\rm total}$: $\rho = -0.11$, and $p$-value $=0.372$).

The diffuse median at $\ell_* \lesssim R_{\rm gal} \lesssim 3\ell_*$ is in agreement with the previously measured diffuse ISM median $R^*_{\rm PAH} = 3.43$ ($\log R^*_{\rm PAH} = 0.54$) (using F2100W instead of F2100W$_{\rm ss}$) for 19/71 of our galaxies by J. Sutter et al. (2024), shown as the black dotted–dashed line in Figures 7 and 8. Although on median the profile of the lowest mass bin ($M_*^{\rm total} \leqslant 10^{9.79} M_\odot$) is in line with the rest of the sample (see Figure 7), in the lowest-$M_*^{\rm total}$ galaxies in our sample with $M_*^{\rm total} \lesssim 10^{9.3}\,M_\odot$, $R^*_{\rm PAH}$ remains universally low in the entire disk relative to more massive galaxies (Figure 8). In particular, our lowest mass galaxy is NGC 3239 (included in Figure 1) with $M_* \approx 10^{9.17}\,M_\odot$, and one of the lowest $\log {\rm R}^*_{\rm PAH} = 0.21$. Table 1 summarizes the median and scatter in $R^*_{\rm PAH}$ across all radial bins for 71 galaxies. Future work integrating our results with JWST observations of lower-mass dwarf galaxies will be useful to test how this trend transitions into the well-known suppression of PAH abundance in low-mass dwarf galaxies (e.g., F. Galliano et al. 2018; A. Li 2020).

In addition to radial profiles, we split each galaxy into distinct morphological and star formation environments (Section 2.3). Table 2 reports the median and spread in mid-

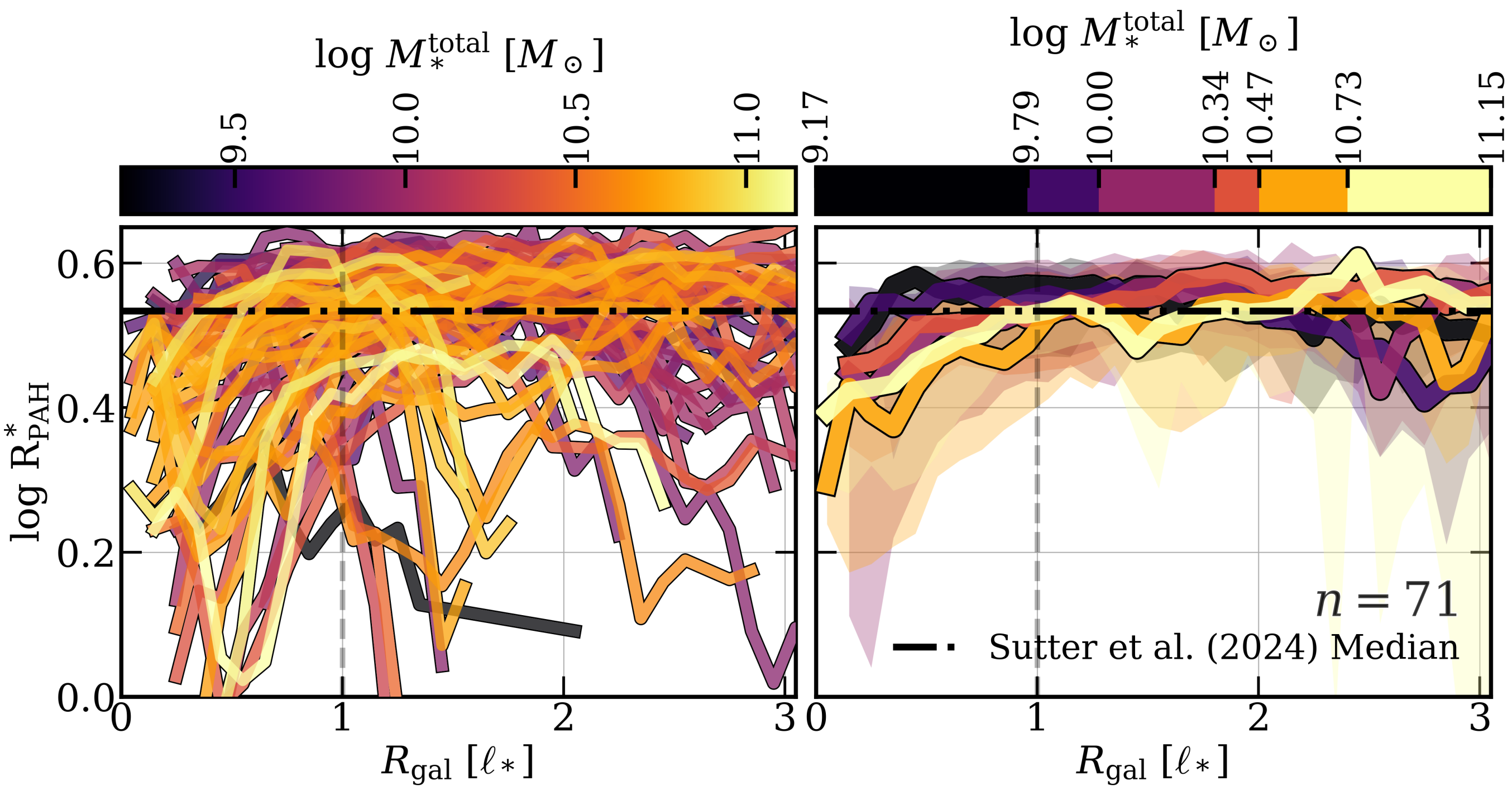

**Figure 7.** Radial profiles of $\log R^*_{\rm PAH}$ for individual galaxies colored by total $M_*$ (left), and then median profiles for groups of galaxies sorted into percentile bins of $M_*$ (right). All Galactocentric distances for the radial profiles are shown in units of the exponential scale length $\ell_*$. For each binned profile, the median profile within a bin (solid line) and 16th–84th percentile range (shaded area) are included. Diffuse ISM median for 19/71 of our targets from J. Sutter et al. (2024) shown as horizontal black dotted–dashed lines.

**Table 1**
Summary of $\log_{10} R^*_{\rm PAH}$ in Diffuse ISM Sight Lines

| Environment | $n$ | Median | Q1/Q3 | $\pm 1\sigma$ |
|---|---|---|---|---|
| $\ell_* \leqslant R_{\rm gal} \leqslant 3\ell_*$ | 71 | 0.55 | 0.49/0.58 | 0.13 |
| | 20 | 0.56 | 0.52/0.59 | 0.11 |
| $R_{\rm gal} < \ell_*$ | 71 | 0.49 | 0.38/0.55 | 0.17 |
| | 20 | 0.47 | 0.35/0.54 | 0.15 |
| Normal Disk | 71 | 0.52 | 0.44/0.56 | 0.11 |
| | 20 | 0.52 | 0.46/0.57 | 0.09 |
| Normal Center | 37/71 | 0.51 | 0.43/0.53 | 0.07 |
| | 8/20 | 0.51 | 0.46/0.52 | 0.05 |
| Young CMZ | 18/71 | 0.39 | 0.31/0.45 | 0.09 |
| | 9/20 | 0.38 | 0.34/0.44 | 0.07 |
| Old Bulge | 13/71 | 0.28 | 0.24/0.33 | 0.13 |
| | 4/20 | 0.31 | 0.28/0.34 | 0.04 |

**Note.** Median, first/third quartiles (Q1/Q3), and standard deviation of the diffuse $\log_{10} R^*_{\rm PAH}$ measured over all annuli within $\ell_* \leqslant R_{\rm gal} \leqslant 3\ell_*$ and $R_{\rm gal} < \ell_*$, and split by galaxy environment, with statistics shown for the full sample $n = 71$ and the subset of $n = 20$ with full MIRI filter coverage.

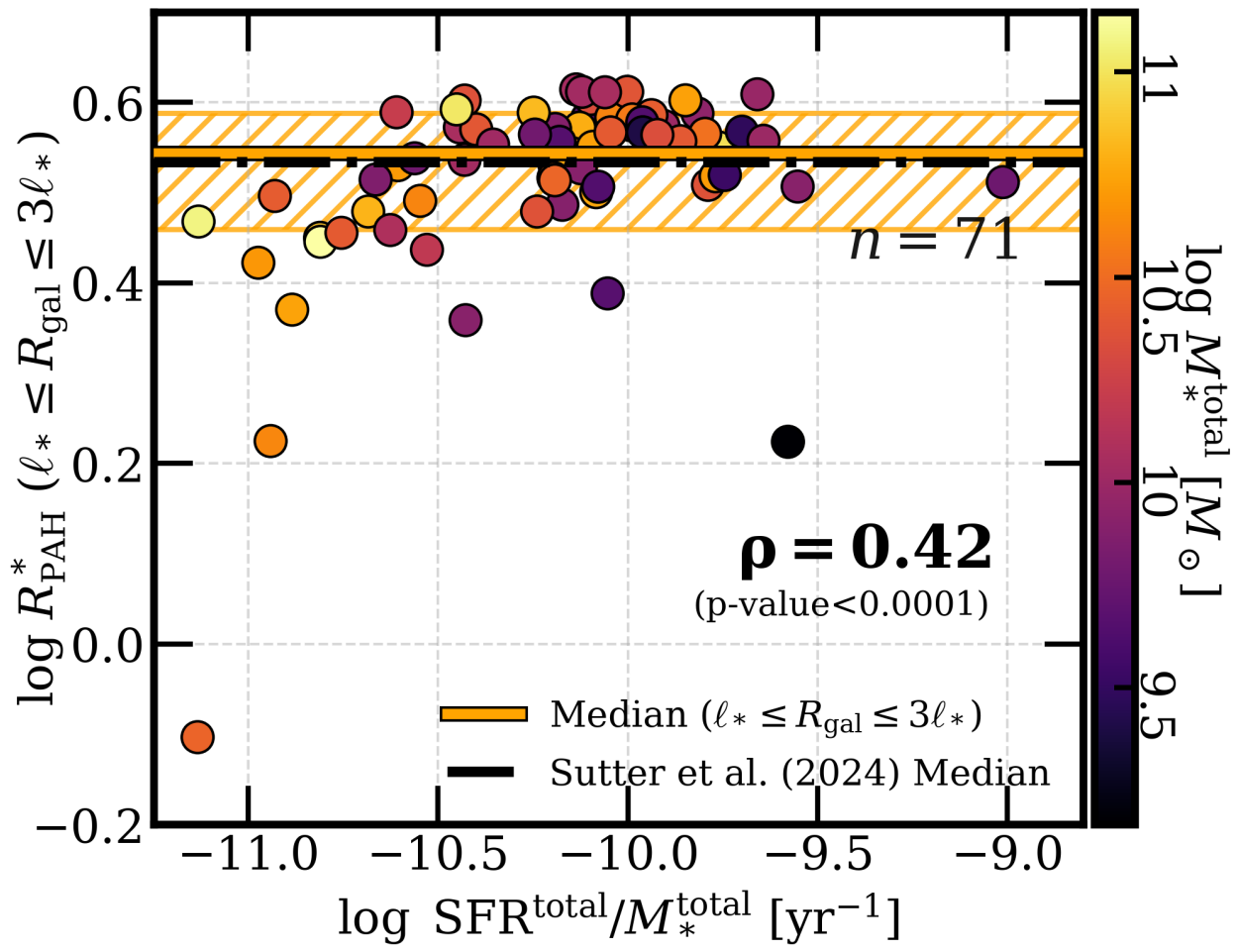

**Figure 8.** Plateau value in radial profiles of $\log R^*_{\rm PAH}$ for each galaxy, measured as the median of all $0.1\ell_*$ annular bins with $\ell_* \leqslant R_{\rm gal} \leqslant 3\ell_*$, points colored by galaxy stellar mass, as a function of global sSFR. Median of 71 galaxies (orange solid line), $1\sigma$ scatter (orange hatched area), diffuse ISM median from J. Sutter et al. (2024, horizontal black dotted–dashed line), and Spearman rank correlation coefficient (printed) shown. See Table 1.

IR colors for each environment,[40] and in Figure 9, we plot $R^*_{\rm PAH}$ and two other key ratios for each environment in each galaxy. We compare these mid-IR colors to the average $\Sigma_{\rm SFR}$, sSFR, and F2100W/F300M color integrated over each environment. We indicate "normal" disk environments with yellow markers and "normal" centers in pale blue (X's if all four MIRI filters are available, circles otherwise).

Across our full sample, most disks show a narrow range of values in $\log R^*_{\rm PAH} \approx 0.52$ (similar to Figure 8), with <0.1 dex scatter (Table 1). Figure 9 also shows a subset of galaxy centers that have similar environment-integrated $\Sigma_{\rm SFR}$ and sSFR to normal disks and also show similar mid-IR colors. "Normal" disks and "normal" centers also show a narrow range of values around $\log {\rm F770W_{ss}/F1130W_{ss}} \approx -0.25$ and

[40] In Table 3 described in the Appendix, we provide the measurements for each environment in each galaxy.

**Table 2**
Summary of Diffuse Mid-IR Colors

| Environment | Median | Q1/Q3 | $\pm 1\sigma$ |
|---|---|---|---|
| $\log_{10}$ F770W$_{ss}$/F1130W$_{ss}$ | | | |
| Normal Disk (20) | −0.24 | −0.28/ − 0.21 | 0.08 |
| Normal Center (8) | −0.26 | −0.28/ − 0.25 | 0.05 |
| Young CMZ (9) | −0.18 | −0.21/ − 0.13 | 0.05 |
| Old Bulge (4) | −0.43 | −0.47/ − 0.40 | 0.04 |
| $\log_{10}$ F1000W$_{ss}$/F2100W$_{ss}$ | | | |
| Normal Disk (20) | −0.24 | −0.26/ − 0.20 | 0.06 |
| Normal Center (8) | −0.19 | −0.21/ − 0.17 | 0.06 |
| Young CMZ (9) | −0.40 | −0.50/ − 0.35 | 0.10 |
| Old Bulge (4) | −0.17 | −0.21/ − 0.14 | 0.04 |
| $\log_{10}$ F770W$_{ss}$/F2100W$_{ss}$ | | | |
| Normal Disk (20) | 0.11 | 0.05/0.16 | 0.09 |
| Normal Center (8) | 0.10 | 0.05/0.11 | 0.05 |
| Young CMZ (9) | −0.03 | −0.07/0.03 | 0.07 |
| Old Bulge (4) | −0.10 | −0.13/ − 0.07 | 0.04 |
| $\log_{10}$ F1130W$_{ss}$/F2100W$_{ss}$ | | | |
| Normal Disk (20) | 0.35 | 0.32/0.38 | 0.05 |
| Normal Center (8) | 0.36 | 0.27/0.38 | 0.08 |
| Young CMZ (9) | 0.08 | 0.01/0.20 | 0.11 |
| Old Bulge (4) | 0.34 | 0.31/0.37 | 0.05 |

**Note.** Median, first/third quartiles, and standard deviation of mid-IR colors in the diffuse ISM of the $n = 20$ galaxies with full MIRI filter coverage (Section 2.1), split by galaxy environment (Section 2.3).

log F1000W$_{ss}$/F2100W$_{ss}$ $\approx -0.20$, with environment-integrated medians and scatter summarized in Table 2.

As with the radial profiles (Figure 8), Figure 9 shows a moderate correlation between environment-integrated sSFR and $R^*_{\rm PAH}$ in "normal" disks ($\rho = 0.47$). We also find a strong correlation between F770W$_{ss}$/F1130W$_{ss}$ and sSFR ($\rho = 0.86$), or F2100W/F300M ($\rho = 0.84$), which we use as a proxy for sSFR (Section 2.4), with best-fit relations reported in Figure 9.

### 3.3. Low $R^*_{\rm PAH}$ in the Centers of Massive Galaxy

Figure 7 shows that $R^*_{\rm PAH}$ declines in the centers of massive galaxies. The most massive galaxies ($M_* \gtrsim 10^{10} M_\odot$) show kiloparsec-scale low $R^*_{\rm PAH}$. Particularly within the central $\sim 1\ell_*$ (typically $\sim$3 kpc), the depression in $R^*_{\rm PAH}$ is significant, 0.1–0.4 dex lower $R^*_{\rm PAH}$ relative to the rest of the disk (Figure 7). Examples include the central regions of NGC 1300 (Figure 3), NGC 1512, and NGC 2775 (Figure 1).

As discussed in Section 2.5.4, we expect uncertainties associated with starlight subtraction to dominate in galaxy centers with faint PAH emission. Figure 9 shows our best-estimate error bars accounting for these uncertainties. The figure shows that the $R^*_{\rm PAH}$ contrasts between (i) bulges and disks and (ii) young centers and disks are both larger than the uncertainty associated with starlight subtraction.

Local conditions in inner galaxies can diverge from those in normal star-forming disks in different ways. CMZs are expected to have increased radiation field intensity because of the high density of young, massive stars. This results in a higher than normal density of H II regions where ionizing photons may destroy fragile PAHs (e.g., J. Sutter et al. 2024; O. V. Egorov et al. 2025, for $R^*_{\rm PAH}$ in H II regions). It also enhances the 21 $\mu$m continuum (e.g., G. J. Bendo et al. 2008). CMZs can also exhibit higher average dust attenuation ($1 \lesssim A_V \lesssim 3$ mag; J. D. T. Smith et al. 2007, see Section 4.5), total gas column (M. Querejeta et al. 2021), and stellar mass density than disks.

The centers of massive galaxies can also host quiescent old stellar population-dominated bulges and bars, where the radiation field is softer, i.e., FUV-poor, due to the low density of young stars and high density of older stars (e.g., the FUV-poor bulge of M31; B. Groves et al. 2012; B. T. Draine et al. 2021). These conditions may be more similar to ETGs and ellipticals (e.g., H. Kaneda et al. 2005; R. Rampazzo et al. 2013). Bulges can also be relatively ISM-free regions where the total gas column density and dust content can be lower than in normal star-forming disks.

Finally, though we have masked the brightest cases (Section 2.5.3), 5 of our galaxies also host active galactic nuclei (AGN; M.-P. Véron-Cetty & P. Véron 2010) which can change PAH and mid-IR continuum emission properties (J. D. T. Smith et al. 2007; J. J. Jensen et al. 2017) over the central $\sim$500 pc, the exact extent of the zone of influence around AGN remains poorly constrained.

These changes in radiation field intensity and shape can change $R^*_{\rm PAH}$ by changing the intensity of the dust continuum, the ionization state or size distribution of PAHs, and/or the importance of stochastic emission versus equilibrium emission from small grains (D. Baron et al. 2025). These, in turn, manifest in the observed mid-IR colors. In Figure 9 and hereafter, we measure these colors for each environment in each target galaxy. We show variations as a function of environment-integrated $\Sigma_{\rm SFR}$ and sSFR. $\Sigma_{\rm SFR}$ correlates with increasing radiation field intensity (e.g., J. Chastenet et al. 2025). It is highest in starburst galaxy centers and lower in normal star-forming disks and bulges. On the other hand, sSFR normalizes the overall star formation surface density by the existing $M_*$, which should correlate with the shape of the radiation field. sSFR is low in FUV-poor bulges and star formation deserts, but higher in normal star-forming disks and galaxy centers.

Figure 9 shows that there is significant variation in diffuse mid-IR colors among environments, and that these mid-IR color variations correlate with $\Sigma_{\rm SFR}$ and sSFR. While $R^*_{\rm PAH}$ remains roughly constant in the diffuse ISM of most normal disks (intermediate $\Sigma_{\rm SFR}$ and sSFR), it decreases in extreme environments—(1) in young CMZs with very high $\Sigma_{\rm SFR}$, and (2) bulges with very low sSFR. Such large variations in local star formation result in changes in both the intensity and shape of the radiation field.

#### 3.3.1. Young Star-forming Centers

The most star-forming centers host CMZs or nuclear stellar disks, where a high density of H II regions leads to widespread PAH destruction, and the overall radiation intensity and dust and gas column densities are higher. These high sSFR centers or CMZs, indicated by dark blue points in Figure 9, show lower $R^*_{\rm PAH}$ than normal disks and normal centers. CMZs on average show 0.15 dex lower $R^*_{\rm PAH}$ than normal star-forming disks, as summarized in Table 1 and Figure 9.

As in Figure 5, in CMZs we show measurements for both the full environment (filled symbols), as well as selecting specifically for only diffuse emission (empty triangles; see Section 2.5.3 for details) for a direct comparison with diffuse emission in disks. This removes from analysis nebular regions in CMZs, where $R^*_{\rm PAH}$ is likely low due to local PAH destruction (J. Sutter et al. 2024; O. V. Egorov et al. 2025), and instead focuses on regions where $R^*_{\rm PAH}$ presumably varies due to physical conditions such as the radiation field, PAH abundance, or PAH properties.

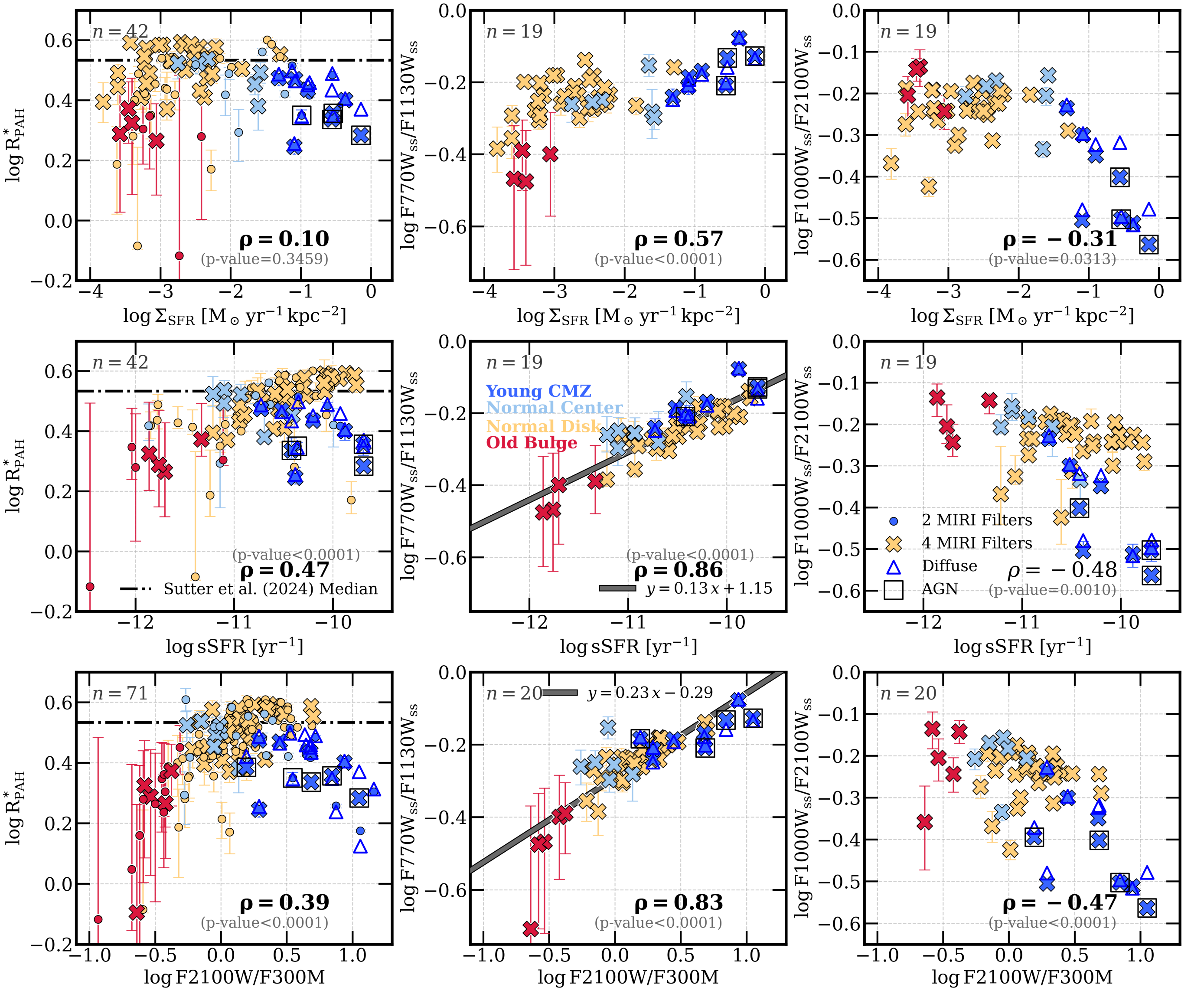

**Figure 9.** $\log R^*_{\rm PAH}$, $\log {\rm F770W_{ss}/F1130W_{ss}}$, and $\log {\rm F1000W_{ss}/F2100W_{ss}}$ as a function of $\log \Sigma_{\rm SFR}$ ($M_\odot$ yr$^{-1}$ kpc$^{-2}$; top) and $\log {\rm sSFR} = \log \Sigma_{\rm SFR}/\Sigma_*$ (yr$^{-1}$; middle) from MUSE, and $\log {\rm F2100W/F300M}$ (bottom row), split by local environment—old stellar bulges (red), young CMZs (bright blue), normal centers (pale blue), and disks (yellow). The markers indicate medians within each environment, with corresponding Spearman rank correlation coefficients ($\rho$) printed. We indicate the 20 galaxies with coverage in all four MIRI filters as X's, and the rest with only two MIRI filters as circles. For galaxy centers, we include medians within the full environment (X's or circles) and select for only diffuse emission (blue triangles), and mark centers with known AGN (open squares; M.-P. Véron-Cetty & P. Véron 2010). The median diffuse $R^*_{\rm PAH}$ from J. Sutter et al. (2024) is included for comparison (black dotted–dashed line).

We summarize the measured median and scatter in diffuse $R^*_{\rm PAH}$ for CMZs in Table 1, $\log R^*_{\rm PAH} \approx 0.39$ with a $\sim$0.1 dex scatter. In the most highly star-forming CMZs ($\Sigma_{\rm SFR} \gtrsim 0.1$ $M_\odot$ yr$^{-1}$ kpc$^{-2}$), diffuse $R^*_{\rm PAH}$ in centers is still $\sim$0.2 dex lower than in "normal" parts of galaxies (Figure 9).

These highly star-forming centers form part of a continuous trend in $\rm F770W_{ss}/F1130W_{ss}$ color versus sSFR (or F2100W/F300M) that includes all environments. At fixed sSFR, diffuse regions in CMZs also show slightly (0.05–0.1 dex) higher $\rm F770W_{ss}/F1130W_{ss}$ than "normal" disks, which could be attributed to harder radiation fields and/or more ionized PAHs in galaxy centers (Section 4.4).

In addition to low $R^*_{\rm PAH}$ and higher $\rm F770W_{ss}/F1130W_{ss}$, CMZs show the lowest $\rm F1000W_{ss}/F2100W_{ss}$ colors compared to all other environments. CMZs show on average $\sim$0.2–0.4 dex lower $\rm F1000W_{ss}/F2100W_{ss}$ compared to all lower $\Sigma_{\rm SFR}$ environments, where $\rm F1000W_{ss}/F2100W_{ss}$ shows a sharp drop at $\Sigma_{\rm SFR} \gtrsim 0.1\,M_\odot$ yr$^{-1}$ kpc$^{-2}$. As we discuss in Section 4.3, the lower $R^*_{\rm PAH}$ and the sharp decline in $\rm F1000W_{ss}/F2100W_{ss}$ in CMZs relative to other environments is consistent with significantly higher radiation field intensities in diffuse parts of the ISM in these galaxy centers.

### *3.3.2. Old Bulges and Star Formation Deserts*

Finally, at the very low end of the sSFR range, $R^*_{\rm PAH}$ also declines. Figure 9 shows that $\log R^*_{\rm PAH}$ is $\sim$0.2–0.6 dex lower relative to "normal" disks in the diffuse ISM of quiescent parts of otherwise star-forming massive galaxies—in bars, bulges, or star formation deserts (which we collectively refer to as "bulges" hereafter). The suppression in $R^*_{\rm PAH}$ in bulges is even lower than that observed in the diffuse ISM of the most highly star-forming CMZs (Table 1). Quiescent bulges show $\log R^*_{\rm PAH} \approx 0.3$, with large ($\gtrsim$0.1 dex) scatter to lower values at the lowest sSFR (or F2100W/F300M).

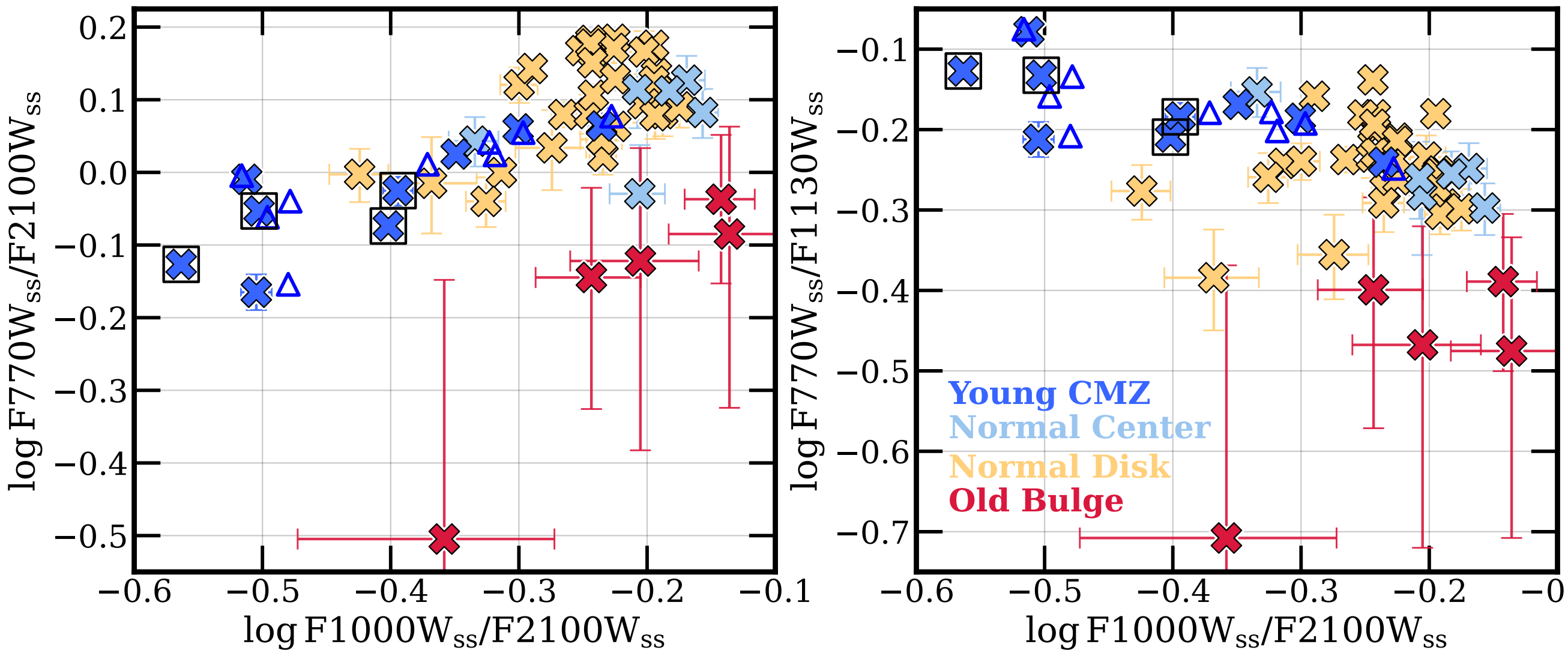

**Figure 10.** Mid-IR color–color variation by environment for 20/71 galaxies with full MIRI filter coverage. F770W$_{ss}$/F2100W$_{ss}$ (or $R^*_{\rm PAH}$) and F770W$_{ss}$/F1130W$_{ss}$ ratios vs. F1000W$_{ss}$/F2100W$_{ss}$ for bulges (red), CMZs (dark blue), "normal" centers (pale blue), and disks (yellow) shown.

Since the bulges are relatively fainter in F770W$_{ss}$ and dominated by strong old stellar emission (bright in F300M), these are the environments where uncertainties in starlight subtraction dominate the error (especially for shorter-wavelength filters such as F770W and F1000W). Even after accounting for these significant uncertainties (added in quadrature to other instrumental and background uncertainties, see Section 2.5.4), included in Figure 9, quiescent bulges show significantly lower $R^*_{\rm PAH}$.

In addition to low $R^*_{\rm PAH}$, bulges also show low $\log {\rm F770W_{ss}/F1130W_{ss}} \approx -0.43$, on average 0.2 dex lower than "normal" disks and centers, and 0.3 dex lower than highly star-forming CMZs. In fact, the 4/13 bulges with F1130W coverage are also in agreement with the tight correlation between F770W$_{ss}$/F1130W$_{ss}$ and sSFR observed across all environments in 20 galaxies. Finally, the 4 bulges with F1000W coverage show F1000W$_{ss}$/F2100W$_{ss}$ colors consistent with in "normal" star-forming parts of galaxies.

The low $R^*_{\rm PAH}$ in regions of low sSFR likely reflects a suppression of 7.7 $\mu$m PAH emission instead of enhanced F2100W$_{ss}$, since F770W$_{ss}$/F1130W$_{ss}$ is low, but F1000W$_{ss}$/F2100W$_{ss}$ is not, which we discuss in detail in Section 4.4. We speculate that the behavior in bulges may reflect the same effect seen in Spitzer spectra of ETGs, which show detected 11.3 $\mu$m emission but very little 7.7 $\mu$m emission (e.g., P. Panuzzo et al. 2011; R. Rampazzo et al. 2013, 2014; D. Baron et al. 2025). Changing radiation field properties are likely not a viable explanation for large variations in F770W$_{ss}$/F1130W$_{ss}$, since the wavelengths are sufficiently close that radiation field considerations are quite small. While the lack of energetic FUV photons (low sSFR regions are also expected to be relatively FUV-poor) can suppress shorter-wavelength PAH emission (e.g., suppress 7.7 $\mu$m emission relative to 11.3 $\mu$m; B. T. Draine et al. 2021), changes in PAH ionization and grain size cause the most significant changes in $R^*_{\rm PAH}$ and F770W$_{ss}$/F1130W$_{ss}$, which we disentangle in Section 4.4. Finally, PAH hydrogenation and composition itself could conceivably be varying in these environments, testing which would require spectroscopic follow-up in bulges.

### 3.4. Mid-IR Color–Color Variation

For the 20 galaxies where all four mid-IR filters are available, we show mid-IR colors as a function of one another to connect the trends seen in Figure 9. In Figure 10, we show F770W$_{ss}$/F2100W$_{ss}$ (or $R^*_{\rm PAH}$) and F770W$_{ss}$/F1130W$_{ss}$ PAH-correlated ratios against the mid-IR continuum ratio F1000W$_{ss}$/F2100W$_{ss}$. Comparing F770W$_{ss}$/F2100W$_{ss}$ and F1000W$_{ss}$/F2100W$_{ss}$, it is clear that in most CMZs and disks, 7.7/21 $\mu$m and 10/21 $\mu$m vary together. If F1000W$_{ss}$ is taken to reflect the continuum near F770W$_{ss}$, this indicates that to first order, mid-IR color variations contrasting "normal" disks versus CMZs are likely primarily driven by variations in the radiation field intensity, not changes in PAH properties. We show in Section 4.3 that environments with low 7.7/21 $\mu$m and 10/21 $\mu$m also show higher far-infrared (far-IR) dust temperature (and hence, radiation intensity), supporting this interpretation.

Figure 10 also highlights the bulges as distinct environments that show abnormally low 7.7/21 $\mu$m despite normal-to-high 10/21 $\mu$m. This departure from the otherwise consistent trend between centers and disks indicates that the low F770W$_{ss}$ in bulges is likely *not* due to changes in only radiation field intensity. The right panel of Figure 10 compares F770W$_{ss}$/F1130W$_{ss}$ and F1000W$_{ss}$/F2100W$_{ss}$. This once again highlights the contrast between highly star-forming CMZs with high F770W$_{ss}$/F1130W$_{ss}$, and quiescent bulges with extremely low F770W$_{ss}$/F1130W$_{ss}$, lower than normal disks. In Section 4.4, we confirm that changes in PAH charge and radiation hardness likely suppress 7.7 $\mu$m emission and result in low $R^*_{\rm PAH}$ in quiescent bulges.

## 4. Interpreting Photometric Band Ratios

### 4.1. Stark Environmental Variation within "Normal" Galaxies

The mid-IR colors of the diffuse ISM in star-forming galaxies vary between "normal" galaxy disks, highly star-forming galaxy centers, and quiescent bulges or star formation deserts. Mid-IR colors in the diffuse ISM on average remain relatively constant across large parts of the optical disks of galaxies, with <0.1 dex variation in colors across large samples of galaxies (Table 2).

Highly star-forming CMZs and quiescent bulges both harbor conditions very different from normal star-forming disks. In CMZs, the average gas and dust column density, and hence $A_V$, can be high. In addition, the high density of young, massive stars leads to a high radiation field intensity and can also result in more ionizing photons that can destroy PAHs and result in harder (FUV-rich) radiation fields (e.g., L. E. Tacconi-Garman et al. 2005; O. V. Egorov et al. 2025), which can enhance dust continuum emission, suppress PAH emission where PAHs are destroyed, or may enhance emission from more ionized, and/or smaller, and/or shorter-wavelength PAH emission. In Section 4.3, we confirm that the variation in mid-IR colors in CMZs is likely driven by changes in the overall radiation field intensity in the diffuse ISM.

On the other hand, quiescent bulges, bars, and star formation deserts are characterized by high stellar surface densities, but can be relatively ISM-free. Although the radiation field intensity can be moderately high in bulges, the radiation spectrum is expected to be FUV-poor (i.e., softer) due to older starlight (e.g., the high radiation intensity but soft, FUV-poor bulge of M31; B. Groves et al. 2012). Softer radiation can suppress short-wavelength PAH emission relative to longer-wavelength features (e.g., B. T. Draine et al. 2021), and may suppress emission from more ionized PAHs. In Section 4.4, we discuss how mid-IR color variations in relatively quiescent bulges may be due to the presence of more neutral PAHs and/or a softer radiation spectrum.

### 4.2. Comparison with Dust Models

In addition to the actual abundance of PAHs, $q_{\mathrm{PAH}}$, the physical properties of PAHs or the properties of the illuminating radiation field (or a combination of the two) may drive changes in mid-IR colors. In the rest of the paper, we try to constrain how much of the variation in mid-IR colors between environments can be explained by local conditions such as changes in the radiation field (intensity and spectrum) and/or PAH properties (charge and size), rather than $q_{\mathrm{PAH}}$. B. T. Draine et al. (2021, hereafter D21) model the effect of each of these factors, and we compare D21 dust model predictions and our mid-IR photometric filter ratios. We use a fixed Milky Way $q_{\mathrm{PAH}}$ throughout, reflecting the relatively high metallicity of our targets, to highlight the range of variation possible due to radiation field and PAH properties only. We compare against three radiation field hardness models from D21 which roughly correspond to the three environments we classify (Section 2.3)—a 3 Myr old (young) starburst (blue; labeled as SB3), a modified Mathis, Mezger, and Panagia (mMMP) radiation field (J. S. Mathis et al. 1983) that roughly corresponds to a Milky Way disk average (yellow; mMMP), and the quiescent Andromeda bulge (red; M31 bulge). These radiation models have decreasing radiation hardness, from most FUV photon-rich (SB3) to most FUV photon-poor (M31 bulge). We note that we use the D21 models directly and produce grids (variation due to changing PAH charge and size) for a given $U$, instead of a range of $U$. While a range of $U$ is likely a more realistic assumption for dust on larger scales, lacking finer constraints, we compare our observations against a range of different (single) $U$ values. We show D21 predictions for the fiducial diffuse ISM $\log U = 1$ models for comparison, and include for reference the predictions for Astrodust+100% neutral PAHs where relevant.

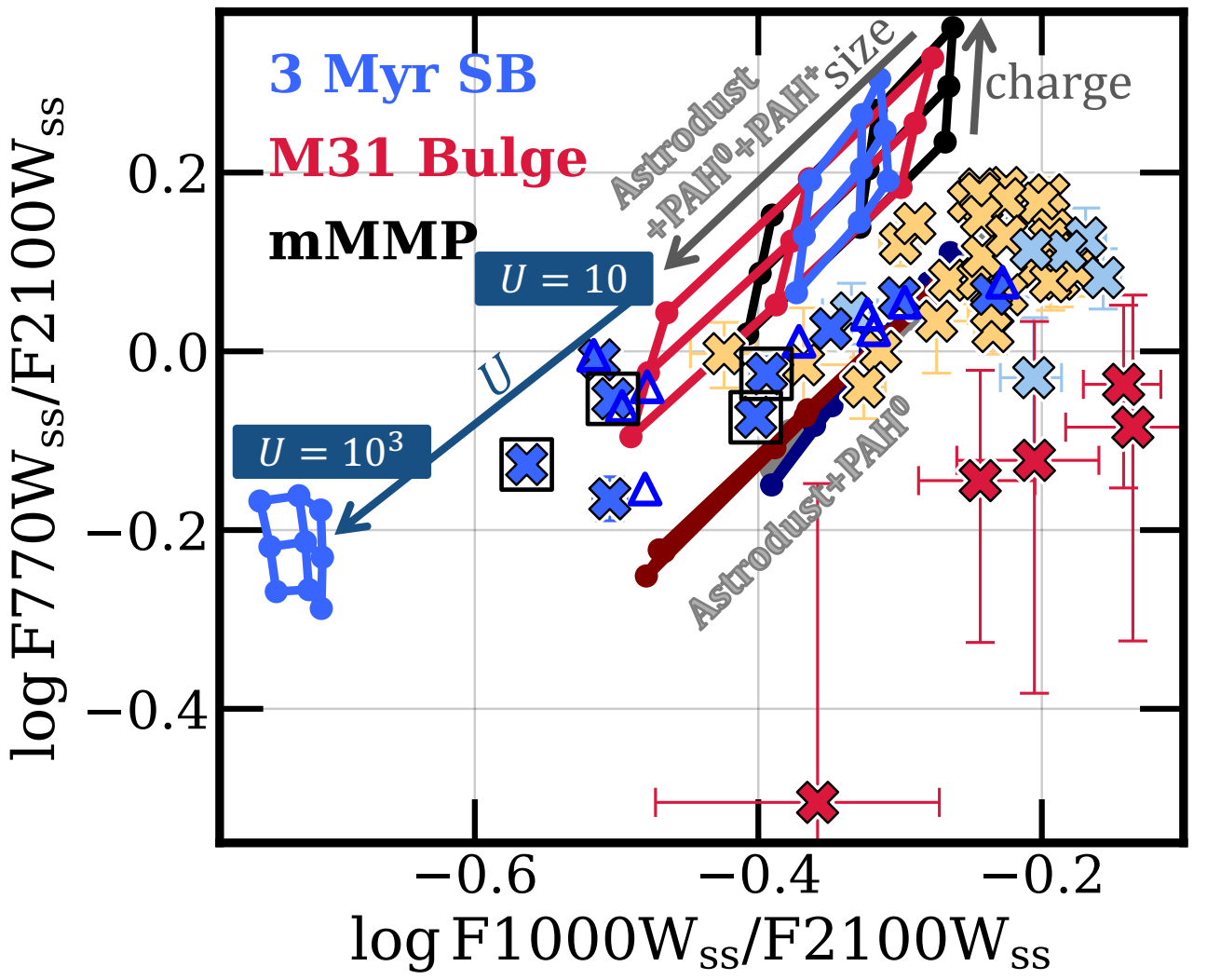

**Figure 11.** The D21 dust model predictions for varying PAH charge and sizes ("grids") assuming three different radiation field hardness models at $\log U = 1$—mMMP (black), M31 bulge (red), and a 3 Myr old starburst (blue); and a 3 Myr old starburst at $\log U = 3$. Arrows show the direction in which grids shift due to increasing PAH ionization, size, and $\log U$. Modulo a 0.1 dex offset in F1000W between model grids and observations, changes in mid-IR colors for CMZs relative to disks are consistent with increasing $U$, while star formation deserts or bulges lie off the standard grids. In darker colors (overlapping gray, maroon, and navy) we show the model predictions at the limiting case where PAHs are 100% neutral ($\mathrm{PAH}^0$) for $\log U = 1$.

We perform synthetic photometry (see, e.g., J. Koornneef et al. 1986; K. D. Gordon et al. 2022) using the transmission curves for the respective JWST filters on the D21 model spectral energy distributions (SEDs) for varying radiation field intensity $U$, three radiation hardness models (SB3, mMMP, and M31 bulge), and varying PAH ionization state and grain size distributions. We use the model SEDs that include both PAH emission and dust continuum. This enables consistent comparison with our measurements, where we remove starlight but not dust continuum emission from MIRI filters.

In Figure 11, we include D21 model grids that show the range of color combinations predicted by varying PAH ionization and size for the three radiation hardness models at fixed $\log U = 1$, and include the SB3 grid at $\log U = 3$ to highlight the direction of change with $\log U$. As noted in Section 3.4, the lower $\mathrm{F770W_{ss}/F2100W_{ss}}$ and lower $\mathrm{F1000W_{ss}/F2100W_{ss}}$ in CMZs compared to disks is in good agreement with variation predicted due to increasing $\log U$ in D21.

Figure 11 shows the observations to be offset from the model grid. This is because the $\mathrm{F1000W_{ss}}$ intensities that we measure are systematically higher than those predicted by the D21 models by 0.1 dex. We identify the $\mathrm{F1000W_{ss}}$ filter as the cause of the 0.1 dex discrepancy, because our observed $\mathrm{F770W_{ss}/F2100W_{ss}}$ colors overlap the range predicted by the models, as do the $\mathrm{F770W_{ss}/F1130W_{ss}}$ and $\mathrm{F1130W_{ss}/F2100W_{ss}}$ colors (shown below, see, e.g., Section 4.4). The bulges also show significantly lower $F770W_{ss}/F2100W_{ss}$ relative to all other environments, which are not reproduced by the standard range of PAH charge variation in the D21 models, but are still bracketed by the limiting case of Astrodust+100% neutral PAHs (darker lines labeled $PAH^0$ in Figure 11), as discussed in Section 4.4.

The offset between D21 and our measurements of F1000W persists across all environments in our sample. We speculate that the discrepancy in F1000W may be due to a combination of (a) the adjacent PAH emission features in the models being too narrow, and so not contributing sufficiently to the F1000W filter; or (b) the shape of the continuum emission around 10 $\mu$m being different in the models. The D21 SEDs do not include differential attenuation of the IR emission, e.g., by the 10 $\mu$m silicate absorption feature, so the low F1000W in the models cannot be due to missing silicate absorption. In fact, if we were to account for the silicate absorption in the models, this would further decrease the F1000W flux from the models and hence would make the discrepancy between the models and the observations even larger (detailed SED comparison forthcoming in M. Boquien et al. 2026, in preparation).

### 4.3. High Radiation Intensity in CMZs

At high radiation intensity ($\log U \gtrsim 2$), a component from dust in thermal equilibrium starts to contribute at the longer MIRI wavelengths, leading to a relative increase in 21 $\mu$m emission. Because $R^*_{PAH}$ includes $F2100W_{ss}$, such intense radiation fields can cause variations in mid-IR color that resemble those due to varying PAH abundance. To check whether the low $R^*_{PAH}$ in highly star-forming CMZs is due to high $U$ or changes in $q_{PAH}$, we require an independent estimate of $U$. We use far-IR colors, which capture variations in the dust temperature, $T_{dust}$. In turn, $T_{dust}$ reflects thermal equilibrium between the dust and the ISRF, and so $U$.

For 31 galaxies, we use archival Herschel PACS 70, 100, and 160 $\mu$m data compiled by J. Chastenet et al. (2025) to measure 70/100 $\mu$m, 100/160 $\mu$m, and 70/160 $\mu$m colors. Though available at much lower resolution (~16″) compared to the JWST observations, these far-IR colors are robust tracers of thermal dust temperature ($T_{dust}$), which is not accessible with JWST alone. For a single radiation field of intensity $U$ and a modified blackbody dust SED with emissivity index $\beta = 2$, $T_{dust} \propto U^{1/6}$. However, similar to the mid-IR (see Section 4.2), far-IR colors can depend on the distribution of radiation field intensities, so the interpretation in terms of a single $T_{dust}$ or $U$ is not straightforward, and we simply present the far-IR colors as a tracer of changes in the average radiation field.

In Figure 12, we show that high sSFR centers indeed show high 70/100 $\mu$m, 100/160 $\mu$m, and 70/160 $\mu$m colors, indicating high average $T_{dust}$, and hence significantly higher $U$ relative to "normal" disks. In addition, some bulges also exhibit moderately high far-IR colors, and hence $T_{dust}$, despite their extremely low sSFR. This is likely due to their high concentration of old stars (high F300M intensities), similar to the bulge of M31 with high $T_{dust}$ due to intense but *old* starlight (B. Groves et al. 2012).

We complement this analysis using less direct but higher resolution mid-IR-based tracers of $U$. We expect that $\Sigma_{SFR}$ and $I_{MIR}/\Sigma_{gas}$, particularly $F2100W_{ss}/\Sigma_{gas}$, should correlate with $U$. This follows since, to first order, in the diffuse ISM with sufficiently low $U$ ($U \lesssim 10$),

$$F2100W_{ss} \propto U \times N(\mathrm{H}) \times \frac{\mathrm{D}}{\mathrm{G}}. \quad (3)$$

The smallest molecules, including PAHs, are stochastically heated and in this single-photon limit their emission is linearly proportional to $U$. At very low $U \sim 1$, F2100W similarly traces stochastically heated grains, and is expected to scale linearly with $U$. At intermediate $U$ grains are still stochastically heated, but multi-photon effects can cause the F2100W emission to become nonlinear as a function of $U$, and finally at larger $U \gtrsim 10$, grains have steady-state temperatures that scale roughly as $U^{1/6}$. Even where F2100W emission and the radiation field intensity are not expected to scale linearly, the relation between F2100W and $U$ is monotonic for realistic ISM conditions (B. T. Draine & A. Li 2001, 2007). So for a fixed dust-to-gas ratio D/G, dust continuum-to-gas column, or $F2100W_{ss}/\Sigma_{gas} \approx F2100W_{ss}/\Sigma_{Mol}$ should scale with the radiation field intensity $U$. While there are caveats, like the need for an $\alpha_{CO}$ prescription to measure $\Sigma_{Mol}$, the possibility of D/G changes, and nonlinearity at high enough $U \gtrsim 10$ (where F2100W increases nonlinearly but still monotonically with $U$), we expect $F2100W_{ss}/\Sigma_{Mol}$ to be a good tracer of relative changes in $U$.

In Figure 12, we test how $F2100W_{ss}/\Sigma_{Mol}$ compares to far-IR colors by combining our $F2100W_{ss}$ JWST measurements with molecular gas maps from PHANGS- Atacama Large Millimeter/submillimeter Array (ALMA; A. K. Leroy et al. 2021). We use the native ~1″ resolution broad masked CO(2–1) maps, which allow for more complete flux retrieval in diffuse regions. We convolve $F2100W_{ss}$ to match ALMA resolution, downsample both maps to the resolution of the ALMA beam, and then measure the median $F2100W_{ss}/I_{CO(2-1)}$ ratio in each environment. We note that we fail to reliably detect some gas-poor bulges and faint disks in $I_{CO(2-1)}$, and hence in $\Sigma_{Mol}$, despite having full-galaxy ALMA coverage, and exclude these low-confidence $\Sigma_{Mol}$ regions from Figures 12 and 13. In addition, while the low resolution of Herschel means we cannot distinguish diffuse emission from overall emission for measuring colors, using the higher 1″ resolution $F2100W_{ss}/I_{CO(2-1)}$ allows us to separate diffuse emission (as in Section 2.5.3).

The correlation between $F2100W_{ss}/I_{CO(2-1)}$ and far-IR colors is weak ($\rho \approx 0.25$–$0.4$). However, much of the scatter appears to be driven by variations in the CO-to-$H_2$ conversion factor, $\alpha_{CO}$, especially the contrast between dense, star-forming centers and more diffuse disks. In the second row of Figure 12, we use $\alpha_{CO}^{2-1}$ maps from J. Sun et al. (2025) to translate $I_{CO(2-1)}$ into molecular gas surface density, $\Sigma_{mol}$. These maps are available for 57 of our galaxies, and use the E. Schinnerer & A. K. Leroy (2024) prescription for estimating $\alpha_{CO}^{2-1}$, accounting for variations in excitation, emissivity, and metallicity.

The correlation between all three far-IR colors and $F2100W_{ss}/\Sigma_{Mol}$ appears much stronger than with $F2100W_{ss}/I_{CO(2-1)}$ ($\rho \approx 73$). The strong correlation supports the use of $F2100W_{ss}/\Sigma_{Mol}$ as a $U$ indicator. The improvement with better $\alpha_{CO}$ treatment is also an indirect validation of the $\alpha_{CO}$ maps we use. Finally, we check that the use of $F2100W_{ss}/\Sigma_{Mol}$ as an indicator of $U$ marks an improvement over simply using MUSE $\Sigma_{SFR}$ (which correlates strongly with F2100W, as shown in Figure 2).

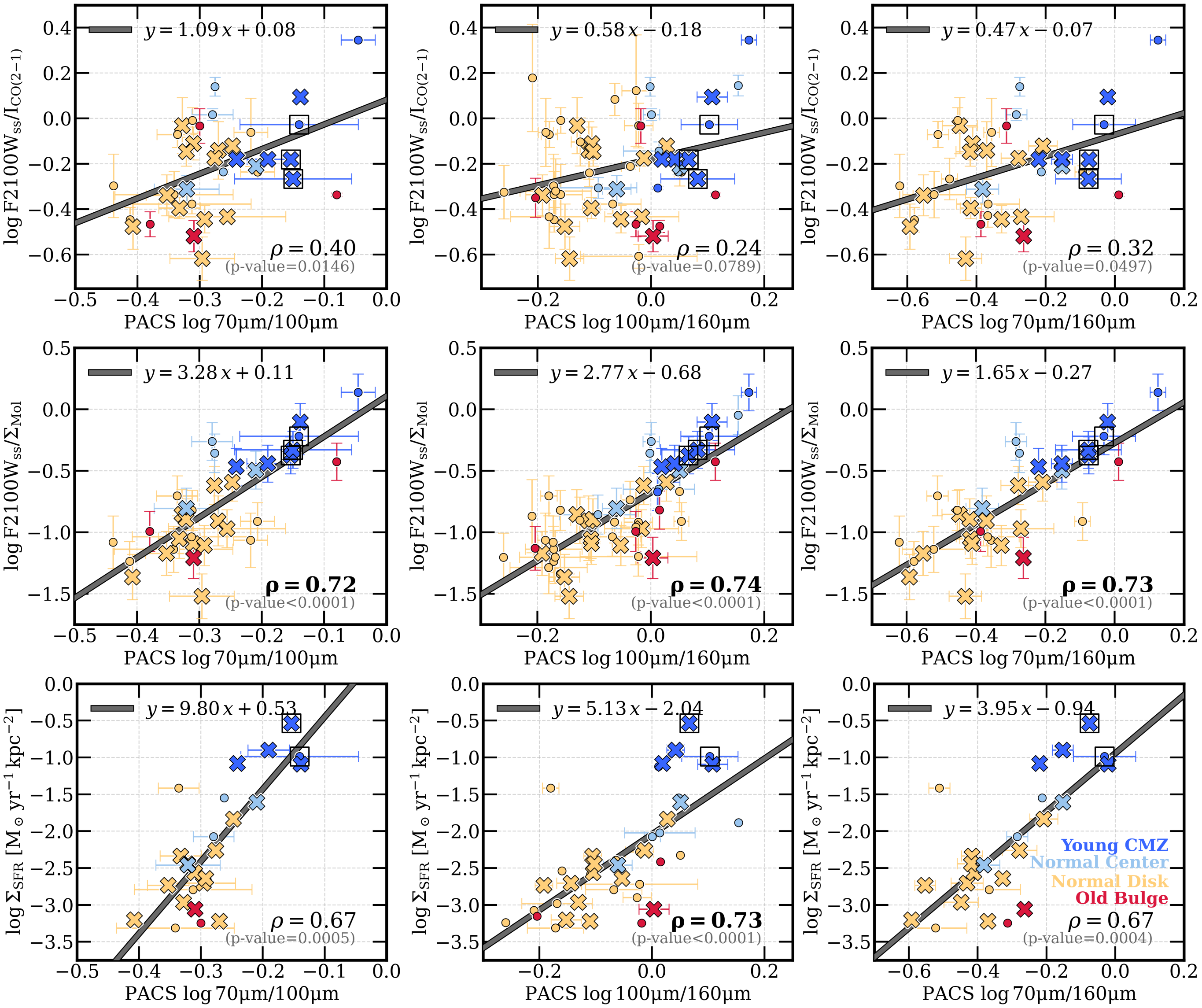

**Figure 12.** Correlating far-IR dust temperature tracers with expected tracers of radiation field intensity: Herschel PACS far-IR ratios of specific intensities ($I_\nu$) for ∼31 galaxies where PACS 70, 100, or 160 $\mu$m data are available, with F2100W$_{ss}$/$I_{CO(2-1)}$ (top row), F2100W$_{ss}$/$\Sigma_{Mol}$ (middle row), and $\Sigma_{SFR}$ (bottom row), data points colored as in previous figures. log−log fits and corresponding Spearman rank correlation coefficients are included for comparison.

We thus interpret the relative changes in F2100W$_{ss}$/$\Sigma_{Mol}$ as relative changes in $U$. Then, from Figure 13, in the diffuse ISM of CMZs, $U$ appears higher than that in most disks. Despite low sSFR, many old stellar bulges also have high $U$. This likely reflects the high intensity of old starlight, e.g., as seen in the bulge of M31 (B. Groves et al. 2012). Finally, most normal star-forming disks show low $U$.

At fixed PAH abundance and high $U$, the apparent $R^*_{PAH}$ depends on $U$ because $U$ affects the dust continuum emission at 7.7 and 21 $\mu$m differently. In Figure 13, we show how our measured $R^*_{PAH}$, F770W$_{ss}$/F1130W$_{ss}$, and F1000W$_{ss}$/F2100W$_{ss}$ vary with F2100W$_{ss}$/$\Sigma_{Mol}$, our rough indicator of $U$, as well as the expected variation in the same colors with varying $U$ for three D21 radiation spectrum models. Comparing the observed trends with F2100W$_{ss}$/$\Sigma_{Mol}$ and the model-predicted trends with $U$, we confirm that the low $R^*_{PAH}$ measured in CMZs is consistent with high local $U$, where F2100W$_{ss}$ rises as a function of $U$, rather than an actual change in PAH mass fraction. The bulges show abnormally low $R^*_{PAH}$ for moderate to low (similar to disks) F2100W$_{ss}$/$\Sigma_{Mol}$, suggesting that the low $R^*_{PAH}$ in bulges is not due to high $U$. This interpretation is further supported by the similarity between the trend in F1000W$_{ss}$/F2100W$_{ss}$ we measure across environments and the model curves for varying $U$. F1000W$_{ss}$/F2100W$_{ss}$ shows a similar trend to $R^*_{PAH}$, with a clear contrast between disks/bulges and CMZs—CMZs show 0.2–0.4 dex lower F1000W$_{ss}$/F2100W$_{ss}$ than all other environments. Hence, the variation in F1000W$_{ss}$/F2100W$_{ss}$ is likely also due to increasing U in CMZs.

Although changes in $R^*_{PAH}$ and F1000W$_{ss}$/F2100W$_{ss}$ in CMZs are thus likely, the correlation between F770W$_{ss}$/F1130W$_{ss}$ and F2100W$_{ss}$/$\Sigma_{Mol}$ is likely not due to changes in $U$. As shown in Figure 13, the D21 models predict that F770W$_{ss}$/F1130W$_{ss}$ is expected to remain relatively unchanged with $U$, and primarily encodes variation in PAH ionization, size, and radiation hardness. Indeed, most of the variation in F770W$_{ss}$/F1130W$_{ss}$ at fixed $U$ in the models reflects changes in ionization. This indicates that diffuse regions in the CMZs may host more ionized, and/or smaller PAHs, and/or harder radiation fields (high F770W$_{\rm ss}$/F1130W$_{\rm ss}$), and bulges likely host larger, and/or less ionized PAHs, and/or softer radiation fields (low F770W$_{\rm ss}$/F1130W$_{\rm ss}$). However, changing radiation field hardness and/or PAH size alone cannot explain the extent of the F770W$_{\rm ss}$/F1130W$_{\rm ss}$ change we measure across environments (spanning roughly 0.5 dex). We further investigate the degeneracies between PAH properties (ionization, size) and radiation hardness in Section 4.4.

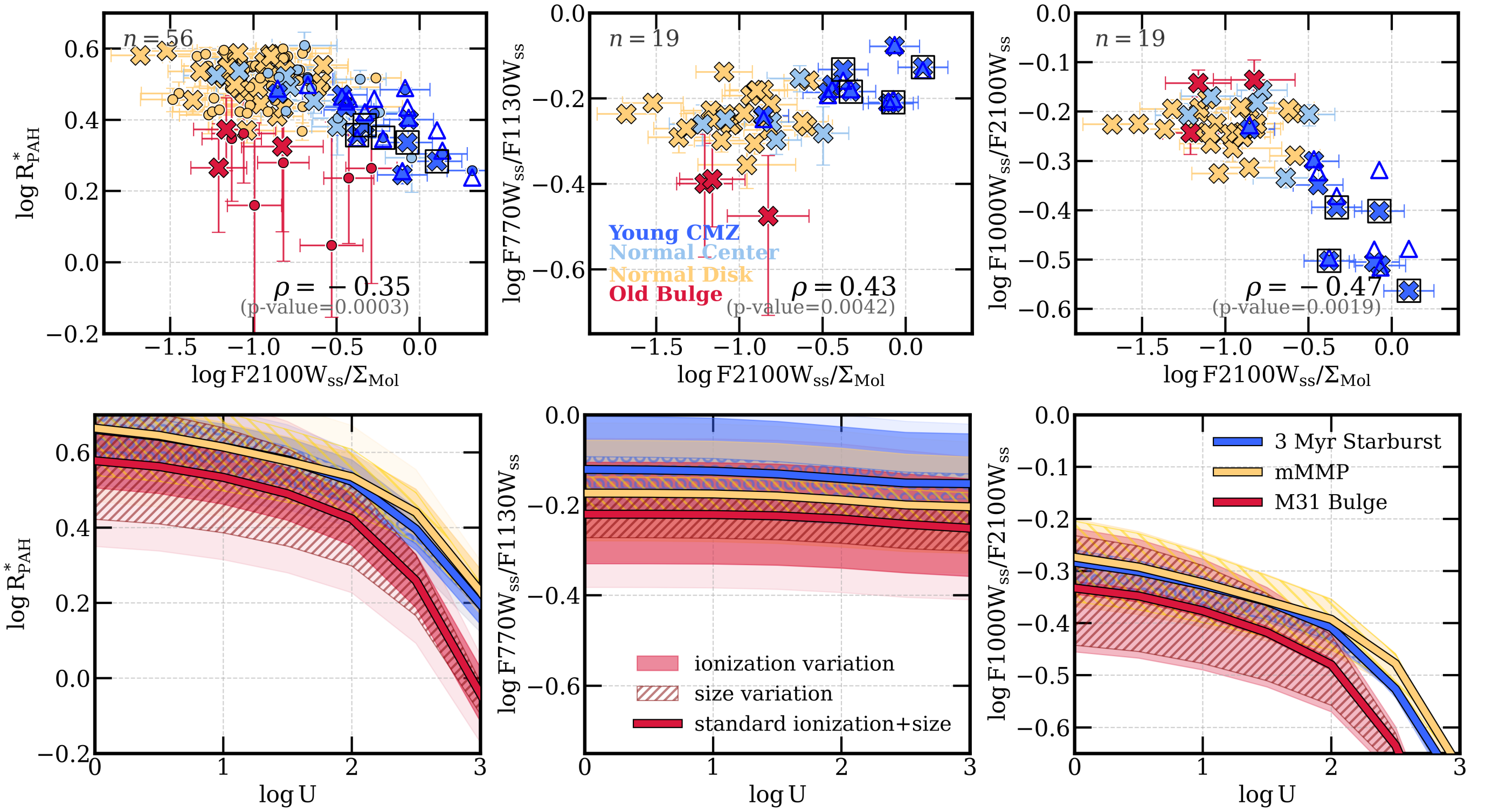

**Figure 13.** Top: $R^*_{\rm PAH}$, F770W$_{\rm ss}$/F1130W$_{\rm ss}$, and F1000W$_{\rm ss}$/F2100W$_{\rm ss}$ as a function of F2100W$_{\rm ss}/\Sigma_{\rm Mol}$, which correlates with $T_{\rm dust}$ (and $U$). Points colored as in previous figures. Bottom: predicted variation in $R^*_{\rm PAH}$, F770W$_{\rm ss}$/F1130W$_{\rm ss}$, and F1000W$_{\rm ss}$/F2100W$_{\rm ss}$ with radiation intensity $\log U$ for three different radiation hardness models from D21—a 3 Myr old starburst (blue), mMMP or Milky Way average (yellow), and the stellar bulge of M31 (red). At each $\log U$, we show the standard value (standard ionization and size; solid lines), maximum variation possible due to PAH ionization at standard grain size (hatches), maximum variation possible due to grain sizes at standard ionization (darker shading), and finally the maximum range of values possible at fixed $U$ from changing both size and ionization in the models (light shading).

### 4.4. PAH Property Variation or Changing Radiation Hardness in Bulges?

In quiescent bulges and star formation deserts, changes in $U$ are not enough to explain observed colors. The bulges show unexpectedly low F770W$_{\rm ss}$ despite low-to-moderate $U$, and normal F1000W$_{\rm ss}$/F2100W$_{\rm ss}$. The low F770W$_{\rm ss}$ despite typical F1000W$_{\rm ss}$ and F2100W$_{\rm ss}$ in bulges can be due to three physical reasons:

1. *Shape of the radiation field spectrum.* PAHs may be illuminated by softer radiation fields in bulges due to the high density of old stars but little to no ongoing star formation. Soft (FUV photon-poor) radiation fields are expected to suppress short-$\lambda$ PAH emission (e.g., D21).
2. *PAH charge.* F770W$_{\rm ss}$ primarily traces emission from the 7.7 $\mu$m PAH feature from more ionized PAHs. The presence of more neutral PAHs thus might explain reduced F770W$_{\rm ss}$/F1130W$_{\rm ss}$ ratios in bulges.
3. *PAH size.* The presence of larger grains[41] can suppress the 7.7 $\mu$m PAH feature relative to 11.3 $\mu$m.

To break this degeneracy between softer radiation, lower PAH ionization, and larger PAH size, we include NIRCam F335M emission available for the 19 Cycle 1 galaxies. We use maps of F335M$_{\rm PAH}$ from Koziol et al. (submitted), which isolate PAH emission from the F335M filter (including aliphatic emission) using a linear combination of the flanking F300M and F360M NIRCam filters to characterize and subtract the relevant stellar continuum, after correcting for any PAH contamination in F360M. Both IR spectroscopy and imaging of metal-rich ISM regions (comparable to our sample) typically exhibit weaker 3.3 $\mu$m PAH emission compared to the longer-wavelength PAH features at 7.7 and 11.3 $\mu$m (e.g., D. A. Dale et al. 2023; A. Maragkoudakis et al. 2026) We thus restrict any comparison that includes F335M$_{\rm PAH}$ (Figure 14, 15) to the subset of pixels within each environment where the fainter F335M$_{\rm PAH}$ is detected at matched 0$''$.9 angular resolution. In lower intensity environments, particularly in bulges and disks, we check that F335M$_{\rm PAH}$ is detected in at least 25% of the pixels in each environment. Hence,

[41] In addition to PAH size, it is important to note that the 7.7/11.3 $\mu$m ratio is also hydrogenation dependent (see B. T. Draine & A. Li 2007, and references therein). For example, in regions where the PAHs have been dehydrogenated, 3.29 and 7.7 $\mu$m emission can be suppressed. However, to first order, we expect the number of H atoms per C atom to depend on the size of the PAH, with higher H/C in smaller PAHs.

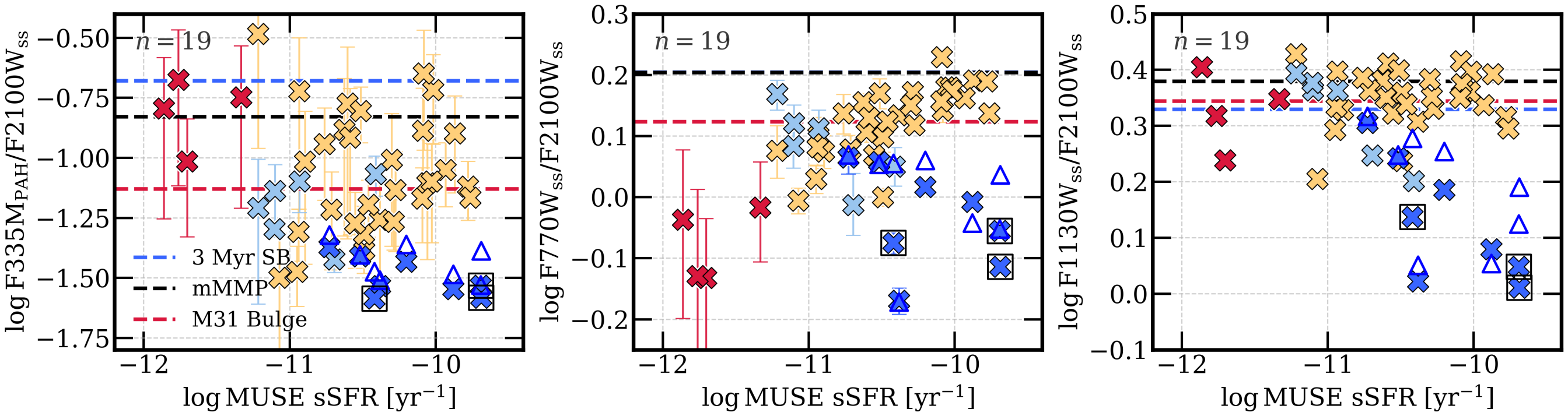

**Figure 14.** PAH-to-continuum ratios for three PAH-dominated filters for the 19 galaxies from Cycle 1, points colored by environment as in previous figures: F335M$_{PAH}$ (from H. Koziol et al. submitted), F770W$_{ss}$, and F1130W$_{ss}$ relative to F2100W$_{ss}$. Dashed lines indicate the predicted variation in respective PAH-to-continuum ratios due to radiation hardness at fixed $\log U = 1$ and standard grain size and ionization from D21.

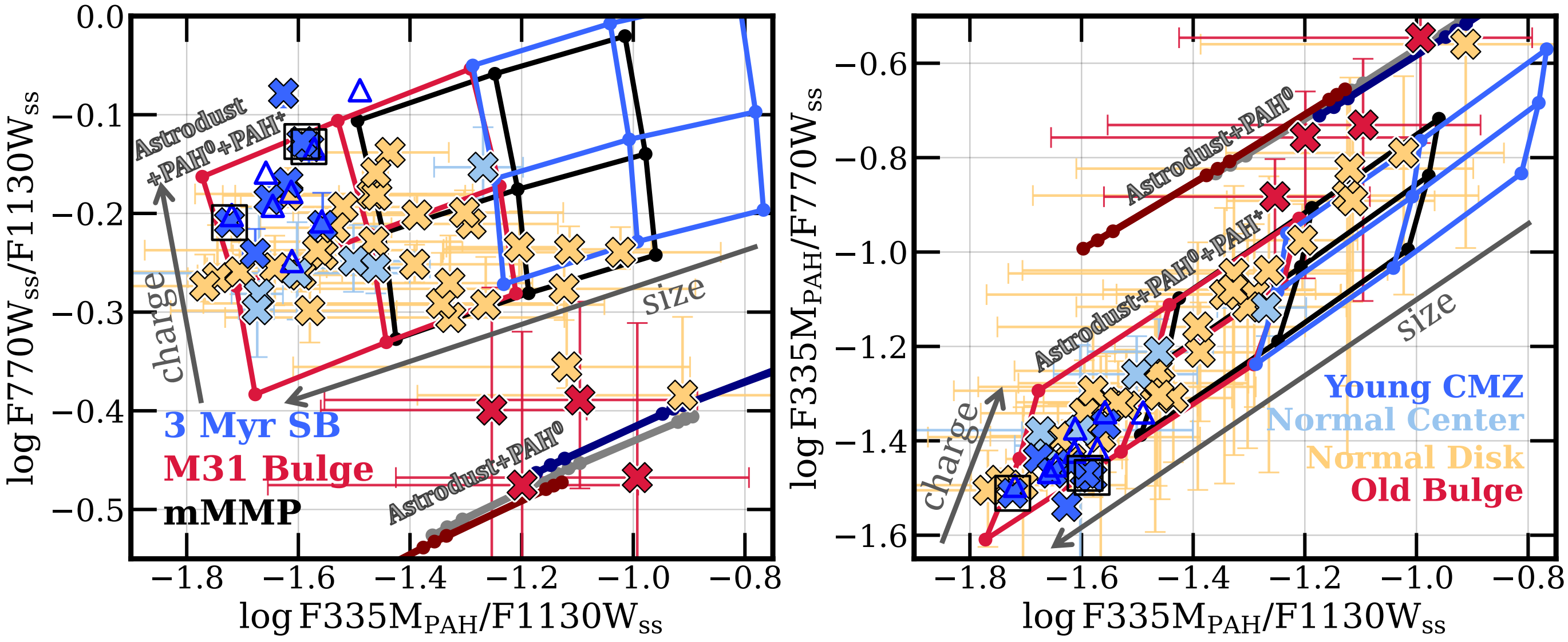

**Figure 15.** Mid-IR and near-IR PAH color–color variation by environment. F770W$_{ss}$/F1130W$_{ss}$ and F335M$_{PAH}$/F770W$_{ss}$ (small, neutral PAH-to-small, ionized PAH) ratios vs. F335M$_{PAH}$/F1130W$_{ss}$ (small, neutral PAH-to-larger, neutral PAH) ratios. D21 dust model grids at $\log U = 1$ included for comparison, with direction of variation due to changing PAH charge and size indicated. No significant variation in dust grids is seen due to varying $\log U$. The darker lines show the limiting case of 100% neutral PAHs (PAH$^0$) at $\log U = 1$ for the three radiation hardness models.

Figures 14 and 15 only include the amount of flux in matched pixels where we reliably detect F335M$_{PAH}$ in addition to the previous MIRI filters. We note that realistic variations in $A_V$ across environments can attenuate at most ~10% and on median <4% of F335M emission, and do not impact our analysis (see Section 4.5).

With the inclusion of F335M$_{PAH}$ we now have access to three PAH-tracing filters—F335M$_{PAH}$ (from small, neutral PAHs), F770W$_{ss}$ (from smaller, ionized PAHs), and F1130W$_{ss}$ (from larger, neutral PAHs). In Figure 14, we present the PAH-to-continuum ratios split by environment for the 19 Cycle 1 galaxies for all three PAH filters, with the predicted filter ratios from D21 for three radiation hardness models at $\log U = 1$ included for comparison (dashed horizontal lines). Once again, all three PAH-to-continuum ratios are low in the CMZs, consistent with higher $U$ and enhanced F2100W$_{ss}$. However, in the bulges, only F770W$_{ss}$/F2100W$_{ss}$ is low, while both F335M$_{PAH}$/F2100W$_{ss}$ and F1130W$_{ss}$/F2100W$_{ss}$ remain high, matching the typical values for normal disks. This rules out softer radiation fields and larger sizes, both of which should also suppress F335M$_{PAH}$/F2100W$_{ss}$. This indicates that the low F770W$_{ss}$ in bulges and star formation deserts is primarily due to PAHs being less ionized. Figure 14 also shows how F770W$_{ss}$/F2100W$_{ss}$ is the PAH-to-continuum ratio that correlates most strongly with local sSFR outside of CMZs, and F1130W$_{ss}$/F2100W$_{ss}$ remains roughly constant across bulges and disks.

Finally, in Figure 15 we compare all environments for the 19 galaxies in PAH color–color space, with the corresponding D21 dust grids (see Section 4.2) at $\log U = 1$ for comparison. Contrasting F770W$_{ss}$/F1130W$_{ss}$, F335M$_{PAH}$/F1130W$_{ss}$, and F335M$_{PAH}$/F770W$_{ss}$ shows that most "normal" galaxy disks and centers scatter within the M31 bulge part of the D21 grids, while CMZs generally show higher F770W$_{ss}$/F1130W$_{ss}$, but still remain within the M31 bulge corner of the grids. Since the D21 models do not include differential attenuation, we speculate that if there is silicate extinction in CMZs, it may move CMZs relative to disks in this direction by primarily attenuating 11.3 $\mu$m. CMZs also have lower F335M$_{PAH}$/F770W$_{ss}$ and F335M$_{PAH}$/F1130W$_{ss}$ than most "normal" star-forming disks, which may indicate the presence of more ionized, larger PAHs in galaxy centers.

Figure 15 once again highlights how quiescent bulges and star formation deserts are unique environments that show anomalously low $F770W_{ss}$, and lie completely outside of the standard parameter space for PAH charge and size variation in the D21 models. Combined with $F335M_{PAH}/F1130W_{ss}$ values (primarily tracing size variation, both from neutral PAHs) in bulges being comparable to many normal disks, this again reinforces the interpretation that the PAH population in bulges is likely more neutral. We note that the colors we measure in bulges are bracketed by the limiting case of the D21 models with 100% neutral PAHs (Astrodust + PAH$^0$; maroon/gray/navy lines in Figure 15). Although a 100% neutral PAH population in star formation deserts is physically unlikely, the better agreement between observed colors in bulges and D21 predictions for neutral PAHs is again consistent with the interpretation that PAHs are likely more neutral in bulges.

Quiescent bulges in our otherwise "normal" star-forming galaxies may present similar ISM conditions as quiescent ETGs and elliptical galaxies, where lower (>500 pc) resolution spectroscopy surveys suggest changes in PAH properties in such extremely FUV-poor environments (e.g., R. Rampazzo et al. 2013). Although softer radiation fields can suppress short-$\lambda$ PAH features including 7.7 $\mu$m, in the M31 bulge, the photometric 7.7 $\mu$m PAH emission is weaker but still detected (B. Groves et al. 2012; B. T. Draine et al. 2014, D21). However, the current range of PAH charge and side distributions included in the D21 model for any radiation spectrum or intensity fails to reproduce the extremely low $F770W_{ss}$ we observe in quiescent bulges in PHANGS galaxies, where much larger variations in PAH ionization may explain the suppressed $F770W_{ss}$ but normal $F1130W_{ss}$. We note that in many ETGs, while the 7.7 $\mu$m PAH feature is virtually undetected, the 11.3 $\mu$m PAH feature is detected (e.g., H. Kaneda et al. 2005; P. Panuzzo et al. 2011; R. Rampazzo et al. 2013, 2014). We speculate that similarly, in extremely low sSFR bulges, the 7.7 $\mu$m PAH feature may be severely suppressed, arguably due to significantly lower PAH charges than what the D21 grids currently allow for, in order to produce the extremely low $F770W_{ss}$ we observe in low sSFR regions.

### 4.5. Other Nondominant Terms

*Moderately high $A_V$ in CMZs.* In addition to high $U$, CMZs and starburst centers are high-density environments where the total gas and dust column density is expected to be higher than in normal disks. In Figure 16, we show that the highest sSFR galaxy centers also host high optical attenuation $A_V$. We measure the ~0.″9-scale $A_V$ using the H$\alpha$/H$\beta$ Balmer decrement and the J. A. Cardelli et al. (1989) extinction curve to estimate $E(B-V)$, and then assume a Milky Way average reddening $R_V = 3.1$. In Figure 16, we present median $A_V$ in environments. Since the medians are measured only at regions where H$\beta$ is reliably detected, allowing for the Balmer decrement measurement, the $A_V$ values in Figure 16 mostly trace the extinction toward the locations of H II regions and their surrounding diffuse ionized gas.

Most highly star-forming galaxy centers show higher attenuation ($A_V \approx$ 1.5–2.5 mag) compared to disks and quiescent bulges ($A_V \approx$ 0.5–1 mag). In high $A_V$ environments, higher silicate absorption at 10 $\mu$m may suppress $F1000W_{ss}$ intensities. In addition, due to the higher $U$ (Section 4.3), CMZs also show elevated $F2100W_{ss}$. So, both high $U$ and higher $A_V$ can contribute to low $F1000W_{ss}/F2100W_{ss}$ in centers (see, e.g., Figure 9). Note that the optical depth $\tau_\lambda = A_\lambda/1.086$, and the observed intensity $I_\lambda^{obs} \propto I_\lambda^{emit} e^{-\tau_\lambda}$. As a result, at $A_V = 1$, roughly 7% and 3.5% of the F1000W and F2100W emission, respectively, is attenuated, and at $A_V = 2$, roughly 14% and 7% of the emission is attenuated (using extinction curves from K. D. Gordon et al. 2023).[42] Even at the shortest wavelength relevant to our analysis (3.36 $\mu$m), roughly 4.5% and 9% of the F335M filter emission is attenuated at $A_V = 1$ and 2, respectively.

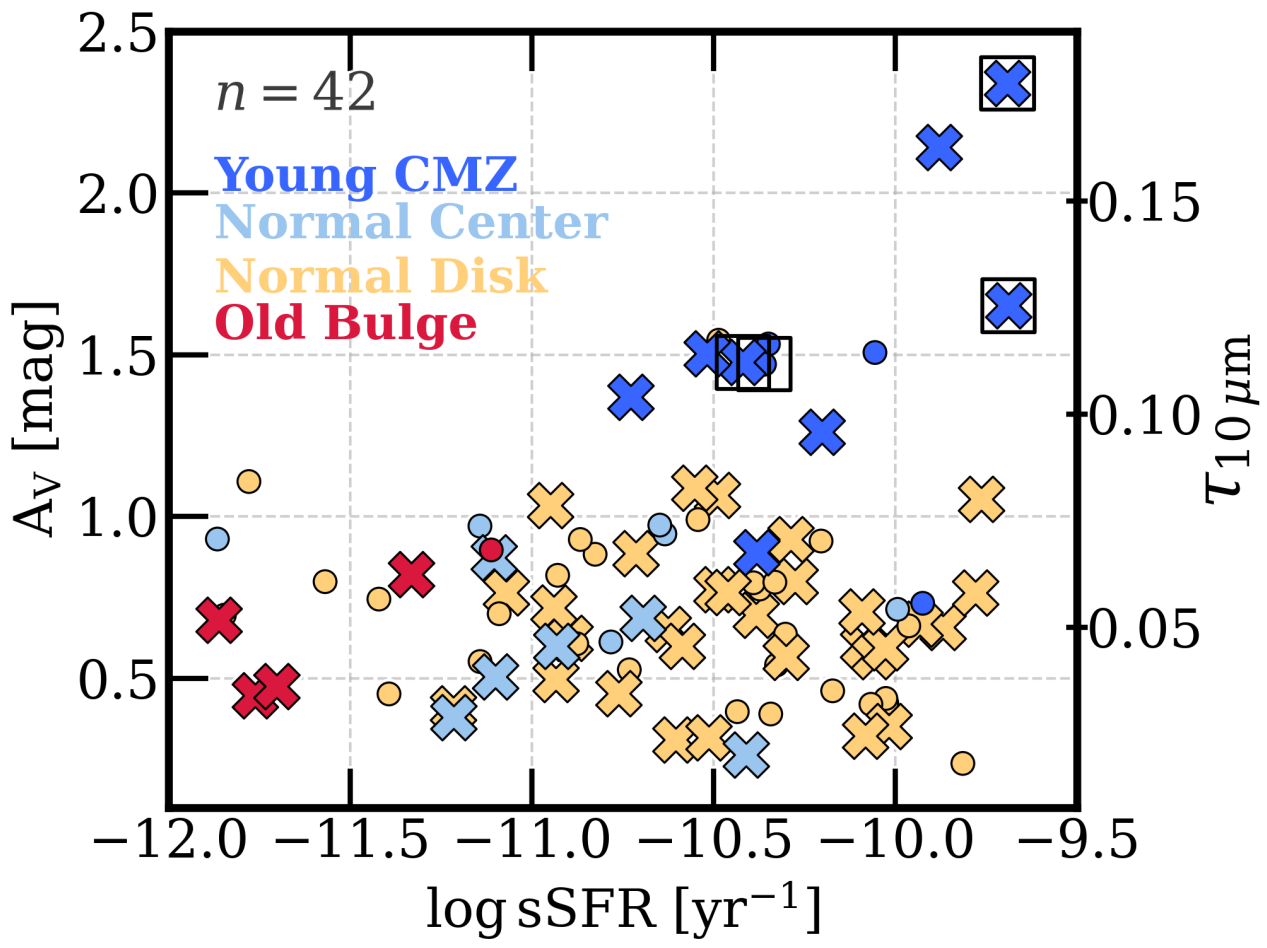

**Figure 16.** MUSE sSFR vs. median (markers) and 16th–84th percentile scatter (error bars) in Balmer decrement attenuation $A_V$ for bulges, centers, and disks in 42 galaxies. Medians and percentiles only measured where both H$\alpha$ and H$\beta$ are detected within each environment. While most low-to-moderate sSFR environments show low $A_V \approx 0.5$, CMZs show higher $A_V$. The alternate $y$-axis shows the translation to $\tau_{10\,\mu m}$, the optical depth at 10 $\mu$m due to silicate absorption (using K. D. Gordon et al. 2023), which in PHANGS CMZs results in <1% attenuation.

The expected decrease due to attenuation is small ($\lesssim$10%) compared to the 0.4 dex decrease in $F1000W_{ss}/F2100W_{ss}$ we observe in CMZs (e.g., Figure 13). Hence, although $A_V$ is higher in highly star-forming centers, $A_V$ is not high enough in PHANGS galaxy centers for silicate absorption to significantly attenuate $F1000W_{ss}$. This is analogous to results from Spitzer spectroscopy for SINGS galaxies, where $A_V$ is rarely high enough ($A_V \gtrsim 5$ mag) for 9.7 $\mu$m silicate absorption to matter (J. D. T. Smith et al. 2007). The consistently low $F1000W_{ss}/F2100W_{ss}$ values in highly star-forming centers of otherwise normal PHANGS galaxies are thus likely due to increasing $F2100W_{ss}$ intensities due to higher average $U$ (Section 4.3). The alternate $y$-axis translates $V$-band attenuation to the corresponding optical depth at 10 $\mu$m due to silicate absorption, assuming the empirical K. D. Gordon et al. (2023) extinction curve at a Milky Way $R_V = 3.1$. For all measured $A_V$ even in CMZs ($A_V \approx$ 1.5–2.5 mag), the relevant optical depth at 10 $\mu$m is ~0.1. So in the CMZs with the highest $A_V$, at most ~15% of the F1000W emission is likely attenuated, which is a negligible effect for our sample.

*AGN.* The presence of strong AGN can also change mid-IR colors due to the changing radiation field and high attenuation (e.g., J. D. T. Smith et al. 2007; L. Zhang & L. C. Ho 2023; G. P. Donnelly et al. 2024). This can include strong continuum emission due to stronger radiation fields, weak PAH emission due to the photodestruction of PAHs, or strong 11.2 and 17 $\mu$m indicating larger PAH sizes closer to the locations of AGN. While several galaxy centers with AGN are removed from our sample due to saturation and strong PSF artifacts in F2100W (see Section 2.5.3), five galaxy centers with AGN are included in our analysis—NGC 1097, NGC 1365, NGC 1672, NGC 4303, and NGC 4826 (M.-P. Véron-Cetty & P. Véron 2010). These AGN show little to no visible signs of mid-IR color variations due to weak AGN, while most strong AGN show strong diffraction artifacts in F2100W, are saturated in galaxy centers, and hence removed from our analysis (see Section 2.5.3). The remaining five AGN galaxy centers show $\lesssim$0.1 dex lower F770W$_{ss}$/F2100W$_{ss}$ (Figure 10), somewhat higher F770W$_{ss}$/F1130W$_{ss}$ (Figure 15), and comparable F335M$_{PAH}$/F1130W$_{ss}$. So in the handful of AGN in our final sample of galaxy centers, we do not recover a statistically significantly change in mid-IR colors due to the presence of AGN.

[42] See also K. D. Gordon et al. 2009; E. L. Fitzpatrick et al. 2019; K. D. Gordon et al. 2021; M. Decleir et al. 2022.

## 5. Summary

Combining high-resolution JWST-MIRI and NIRCam imaging data for 71 nearby star-forming galaxies, we measure the variation in mid-IR colors and observed PAH fraction $R^{*}_{PAH} = 2.57 \times$ F770W$_{ss}$/F2100W$_{ss}$ (Equation (1)) in the diffuse ISM as a function of Galactocentric radius and morphological environment (disks, centers, bars; Section 2.3). We compare these with complementary ionized gas and SFR information from VLT/MUSE, molecular gas information from ALMA, and Herschel far-IR dust colors to investigate how mid-IR colors and PAH band ratios vary as a function of radiation field intensity, radiation field hardness, PAH charge, and PAH size, and optical attenuation across galaxies. We find stark differences in measured mid-IR colors in the diffuse ISM (i.e., regions not dominated by nebular emission) when contrasting "normal" star-forming disks and centers with highly star-forming CMZs, and quiescent old stellar bulges.

1. Mid-IR colors and the observed PAH fraction $R^{*}_{PAH}$ remain fairly constant across diffuse parts of "normal" star-forming disks, with some variation due to global sSFR (Section 3.2, Figure 8, Tables 1, 2). The typical diffuse $\log_{10}R^{*}_{PAH}$ across the disks of 71 nearby galaxies is 0.52, with 0.11 dex standard deviation from galaxy to galaxy.
2. The centers ($R_{gal} < \ell_{*}$, the exponential scale length of the stellar disk) of massive ($M_{*} \gtrsim 10^{10.5}\,M_{\odot}$) star-forming galaxies show ∼0.2–0.4 dex lower $R^{*}_{PAH}$ relative to disks (Figure 7, Table 1). We identify two distinct environments in galaxy centers where $R^{*}_{PAH}$ is suppressed—highly star-forming CMZs, and quiescent bulges or star formation deserts (Section 3.3).
3. The diffuse ISM in highly star-forming CMZs shows low $R^{*}_{PAH}$ and PAH-to-continuum ratios, with median $\log R^{*}_{PAH} = 0.39$ with 0.09 dex scatter, high $\log$ F770W$_{ss}$/F1130W$_{ss} \approx -0.18$, and low $\log$ F1000W$_{ss}$/F2100W$_{ss} \approx -0.40$ (Section 3.3.1, Table 2, Figure 9). The changes in mid-IR colors in CMZs are consistent with higher average radiation intensity $U$ (Section 4.3, Figure 13).
4. We show the efficacy of using F2100W$_{ss}/\Sigma_{Mol}$ as a tracer of average radiation field intensity $U$ in environments, and validate this approach against Herschel far-IR dust colors and MUSE $\Sigma_{SFR}$ (Figure 12).
5. Diffuse ISM sight lines of quiescent bulges and star formation deserts show low $R^{*}_{PAH}$ relative to "normal" disks (Section 3.3.2), with median $\log R^{*}_{PAH} = 0.28$ with ∼0.15 dex scatter (Table 1). For 4/13 bulges, we confirm that F770W$_{ss}$ is the only PAH-correlated filter that is anomalously low in bulges, while other PAH filters (F335M$_{PAH}$ and F1130W$_{ss}$; Figure 14), and the continuum (F1000W$_{ss}$/F2100W$_{ss}$) appear similar to "normal" disks (Table 2, Figure 15). We speculate that bulges likely host diffuse ISM conditions similar to ETGs and elliptical galaxies, and that the extremely low F770W$_{ss}$ in bulges (0.1–0.25 dex lower than disks) is likely due to an underlying PAH population that is predominantly neutral (Section 4.4). The extremely low F770W$_{ss}$/F1130W$_{ss}$ and high F335M$_{PAH}$/F770W$_{ss}$ colors measured in bulges are beyond what current dust models can reproduce (Figure 15; D21).
6. The F1000W$_{ss}$ intensities measured across PHANGS galaxies are systematically $\gtrsim$0.1 dex higher than what current dust models can predict (Figure 11; D21), and the discrepancy between D21 models and observations is largest in bulges and "normal" disks.

### Acknowledgments

We thank the anonymous referee for their constructive comments.

This work has been carried out as part of the PHANGS Collaboration. This work is based on observations made with the NASA/ESA/CSA JWST. The data were obtained from the Mikulski Archive for Space Telescopes at the Space Telescope Science Institute, which is operated by the Association of Universities for Research in Astronomy, Inc., under NASA contract NAS 5-03127 for JWST. These observations are associated with GO programs #2107, #3177, and #3707. Support for programs #2107, #3177, and #3707 was provided by NASA through a grant from the Space Telescope Science Institute, which is operated by the Association of Universities for Research in Astronomy, Inc., under NASA contract NAS 5-03127. The specific JWST observations analyzed can be accessed via DOI: 10.17909/ew88-jt15 and 10.17909/yz9g-pd22. The PHANGS-MUSE LP data can be accessed at ESO DOI: 10.18727/archive/47.

This work is also based on observations collected at the European Southern Observatory under ESO programs 094.C-0623 (PI: Kreckel), 095.C-0473, 098.C-0484 (PI: Blanc), 1100.B-0651 (PHANGS-MUSE; PI: Schinnerer), as well as 094.B-0321 (MAGNUM; PI: Marconi), 099.B-0242, 0100.B-0116, 098.B-0551 (MAD; PI: Carollo) and 097.B-0640 (TIMER; PI: Gadotti).

This paper uses the following ALMA data, which have been processed as part of the PHANGS–ALMA survey: ADS/JAO.ALMA#2012.1.00650.S, ADS/JAO.ALMA#2013.1.00803.S, ADS/JAO.ALMA#2013.1.01161.S, ADS/JAO.ALMA#2015.1.00121.S, ADS/JAO.ALMA#2015.1.00782.S, ADS/JAO.ALMA#2015.1.00925.S, ADS/JAO.ALMA#2015.1.00956.S, ADS/JAO.ALMA#2016.1.00386.S, ADS/JAO.ALMA#2017.1.00392.S, ADS/JAO.ALMA#2017.1.00766.S, ADS/JAO.ALMA#2017.1.00886.L, ADS/JAO.ALMA#2018.1.01321.S, ADS/JAO.ALMA#2018.1.01651.S, ADS/JAO.ALMA#2018.

A.00062.S, ADS/JAO.ALMA#2019.1.01235.S, ADS/JAO.ALMA#2019.2.00129.S. ALMA is a partnership of ESO (representing its member states), NSF (USA) and NINS (Japan), together with NRC (Canada), NSC and ASIAA (Taiwan), and KASI (Republic of Korea), in cooperation with the Republic of Chile. The Joint ALMA Observatory is operated by ESO, AUI/NRAO, and NAOJ. The National Radio Astronomy Observatory is a facility of the National Science Foundation operated under cooperative agreement by Associated Universities, Inc.

D.P. is supported by the NSF GRFP.

A.K.L. and D.P. gratefully acknowledge support from NSF AST AWD 2205628, JWST-GO-02107.009-A, and JWST-GO-03707.001-A. A.K.L. also gratefully acknowledges support by a Humboldt Research Award.

K.S., H.K, and M.P. acknowledge funding support from JWST-GO-2107.006-A and JWST-GO-3707.005-A.

T.G.W. gratefully acknowledges support from the UK ALMA Regional Centre (ARC) Node, which is supported by the Science and Technology Facilities Council grant number ST/Y004108/1.

R.S.K. acknowledges financial support from the ERC via Synergy Grant "ECOGAL" (project ID 855130) and from the German Excellence Strategy via the Heidelberg Cluster "STRUCTURES" (EXC 2181-390900948). In addition, R.S.K. is grateful for funding from the German Ministry for Economic Affairs and Climate Action in project "MAINN" (funding ID 50OO2206), and from DFG and ANR for project "STARCLUSTERS" (funding ID KL 1358/22-1).

J.G.L. acknowledges funding from the DLR (German Aerospace Agency) via grant 50 OR2401.

M.B. acknowledges support by the ANID BASAL project FB210003. This work was supported by the French government through the France 2030 investment plan managed by the National Research Agency (ANR), as part of the Initiative of Excellence of Université Côte d'Azur under reference No. ANR-15-IDEX-01. This research was funded, in whole or in part, by the French National Research Agency (ANR), grant ANR-24-CE92-0044 (project STARCLUSTERS).

*Facilities:* JWST, VLT:Yepun, ALMA, Herschel, WISE.

*Software:* astropy (Astropy Collaboration et al. 2013, 2018).

## Appendix

This appendix presents a machine-readable table that contains our measurements. Table 3 gives the names, units, and a short description of each column in the table.

**Table 3**
Columns in the Value-added Table

| Column | Unit | Description |
|---|---|---|
| galaxy | ⋯ | Galaxy name |
| morphological_environment | ⋯ | Morphological environment following M. Querejeta et al. (2021, Section 2.3) |
| morph_SF_environment | ⋯ | Morphological+SF environment classification (Section 2.3) |
| logF300M | log MJy sr$^{-1}$ | Median F300M intensity (Section 2.1.2) |
| logF2100W | log MJy sr$^{-1}$ | Median F2100W intensity (Section 2.1.2) |
| logF2100Wss | log MJy sr$^{-1}$ | Median F2100W$_{ss}$ intensity (Section 2.1.2) |
| logF770Wss | log MJy sr$^{-1}$ | Median F770W$_{ss}$ intensity (Section 2.1.2) |
| logF1000Wss | log MJy sr$^{-1}$ | Median F1000W$_{ss}$ intensity (Section 2.1.2) |
| logF1130Wss | log MJy sr$^{-1}$ | Median F1130W$_{ss}$ intensity (Section 2.1.2) |
| MUSE_logsSFR | log yr$^{-1}$ | Environment-integrated sSFR (Section 2.2) |
| MUSE_logSigmaSFR | log $M_\odot$ yr$^{-1}$ kpc$^{-2}$ | Environment-integrated $\Sigma_{SFR}$ (Section 2.2) |
| MUSE_Av | mag | Balmer decrement $V$-band attenuation (Section 2.2) |
| logRPAH*ss | ⋯ | Median $R^*_{PAH}$ (Section 2.1.3) |
| logF2100W/F300M | ⋯ | Median F2100W/F300M (Section 2.4) |
| logF770Wss/F2100Wss | ⋯ | Median F770W$_{ss}$/F2100W$_{ss}$ (Section 2.5.2) |
| logF770Wss/F1130Wss | ⋯ | Median F770W$_{ss}$/F1130W$_{ss}$ (Section 2.5.2) |
| logF1000Wss/F2100Wss | ⋯ | Median F1000W$_{ss}$/F2100W$_{ss}$ (Section 2.5.2) |
| logPACS100/160_conv160 | ⋯ | Median 100 $\mu$m/160 $\mu$m (Section 4.3) |
| logPACS70/100_conv100 | ⋯ | Median 70 $\mu$m/100 $\mu$m (Section 4.3) |
| logPACS70/160_conv160 | ⋯ | Median 70 $\mu$m/160 $\mu$m (Section 4.3) |
| logF2100Wss/ICO21 | log MJy sr$^{-1}$/(K km s$^{-1}$) | Median F2100W$_{ss}$/$I_{CO(2-1)}$ (Section 4.3) |
| logF2100Wss/SigmaMol | log MJy sr$^{-1}$/($M_\odot$pc$^{-2}$) | Median F2100W$_{ss}$/$\Sigma_{Mol}$ (Section 4.3) |
| logF335MPAH/F770Wss | ⋯ | Median F335M$_{PAH}$/F770W$_{ss}$ (Section 4.4) |
| logF335MPAH/F1130Wss | ⋯ | Median F335M$_{PAH}$/F1130W$_{ss}$ (Section 4.4) |

**Note.** This table introduces the environment-integrated measurements made available in machine-readable form, with the basic environment properties needed to reproduce most figures.

(This table is available in its entirety in machine-readable form in the online article.)

## ORCID iDs

Debosmita Pathak https://orcid.org/0000-0003-2721-487X
Karin M. Sandstrom https://orcid.org/0000-0002-4378-8534
Adam K. Leroy https://orcid.org/0000-0002-2545-1700
Ryan Chown https://orcid.org/0000-0001-8241-7704
Daniel A. Dale https://orcid.org/0000-0002-5782-9093
Hannah B. Koziol https://orcid.org/0009-0001-5949-1524
Mansi Padave https://orcid.org/0000-0002-3472-0490
Jessica Sutter https://orcid.org/0000-0002-9183-8102
Thomas G. Williams https://orcid.org/0000-0002-0012-2142
Dalya Baron https://orcid.org/0000-0003-4974-3481
Alberto D. Bolatto https://orcid.org/0000-0002-5480-5686
Médéric Boquien https://orcid.org/0000-0003-0946-6176
Oleg V. Egorov https://orcid.org/0000-0002-4755-118X
Eric Emsellem https://orcid.org/0000-0002-6155-7166
Simon C. O. Glover https://orcid.org/0000-0001-6708-1317
Ralf S. Klessen https://orcid.org/0000-0002-0560-3172
Eva Schinnerer https://orcid.org/0000-0002-3933-7677
Todd A. Thompson https://orcid.org/0000-0003-2377-9574
Francesco Belfiore https://orcid.org/0000-0002-2545-5752
Frank Bigiel https://orcid.org/0000-0003-0166-9745
Enrico Congiu https://orcid.org/0000-0002-8549-4083
Ivan Gerasimov https://orcid.org/0000-0001-7113-8152
Jay González Lobos https://orcid.org/0000-0002-6056-3425
Annie Hughes https://orcid.org/0000-0002-9181-1161
Laura A. Lopez https://orcid.org/0000-0002-1790-3148
Justus Neumann https://orcid.org/0000-0002-3289-8914
Elias K. Oakes https://orcid.org/0000-0002-0119-1115
Hsi-An Pan https://orcid.org/0000-0002-1370-6964
Erik Rosolowsky https://orcid.org/0000-0002-5204-2259
Sumit K. Sarbadhicary https://orcid.org/0000-0002-4781-7291

## References

Allamandola, L. J., Hudgins, D. M., & Sandford, S. A. 1999, ApJL, 511, L115
Allamandola, L. J., Tielens, A. G. G. M., & Barker, J. R. 1989, ApJS, 71, 733
Aniano, G., Draine, B. T., Hunt, L. K., et al. 2020, ApJ, 889, 150
Astropy Collaboration, Price-Whelan, A. M., Sipőcz, B. M., et al. 2018, AJ, 156, 123
Astropy Collaboration, Robitaille, T. P., Tollerud, E. J., et al. 2013, A&A, 558, A33
Bakes, E. L. O., & Tielens, A. G. G. M. 1994, ApJ, 427, 822
Barnes, A. T., Chandar, R., Kreckel, K., et al. 2026, A&A, 706, A95
Baron, D., Sandstrom, K. M., Rosolowsky, E., et al. 2024, ApJ, 968, 24
Baron, D., Sandstrom, K. M., Sutter, J., et al. 2025, ApJ, 978, 135
Belfiore, F., Leroy, A. K., Sun, J., et al. 2023a, A&A, 670, A67
Belfiore, F., Leroy, A. K., Williams, T. G., et al. 2023b, A&A, 678, A129
Bendo, G. J., Draine, B. T., Engelbracht, C. W., et al. 2008, MNRAS, 389, 629
Boquien, M., Burgarella, D., Roehlly, Y., et al. 2019, A&A, 622, A103
Burkhardt, A. M., Long Kelvin Lee, K., Bryan Changala, P., et al. 2021, ApJL, 913, L18
Bushouse, H., Eisenhamer, J., Dencheva, N., et al. 2025, JWST Calibration Pipeline, v1.19.1, Zenodo, doi:10.5281/zenodo.6984365
Calzetti, D., Kennicutt, R. C., Adamo, A., et al. 2025, ApJ, 991, 198
Cardelli, J. A., Clayton, G. C., & Mathis, J. S. 1989, ApJ, 345, 245
Chastenet, J., Sandstrom, K., Leroy, A. K., et al. 2025, ApJS, 276, 2
Chastenet, J., Sutter, J., Sandstrom, K., et al. 2023, ApJL, 944, L12
Chown, R., Leroy, A. K., Sandstrom, K., et al. 2025, ApJ, 983, 64
Chown, R., Okada, Y., Peeters, E., et al. 2025c, A&A, 698, A86
Dale, D. A., Bendo, G. J., Engelbracht, C. W., et al. 2005, ApJ, 633, 857
Dale, D. A., Boquien, M., Turner, J. A., et al. 2023, AJ, 165, 260
Dale, D. A., Cohen, S. A., Johnson, L. C., et al. 2009, ApJ, 703, 517
Dale, D. A., Graham, G. B., Barnes, A. T., et al. 2025, AJ, 169, 133
Decleir, M., Gordon, K. D., Andrews, J. E., et al. 2022, ApJ, 930, 15
Donnelly, G. P., Smith, J. D. T., Draine, B. T., et al. 2024, ApJ, 965, 75
Draine, B. T., & Li, A. 2001, ApJ, 551, 807
Draine, B. T., & Li, A. 2007, ApJ, 657, 810
Draine, B. T., Li, A., Hensley, B. S., et al. 2021, ApJ, 917, 3
Draine, B. T., Aniano, G., Krause, O., et al. 2014, ApJ, 780, 172
Draine, B. T., Dale, D. A., Bendo, G., et al. 2007, ApJ, 663, 866
Egorov, O. V., Leroy, A. K., Sandstrom, K., et al. 2025, A&A, 703, A103
Emsellem, E., Schinnerer, E., Santoro, F., et al. 2022, A&A, 659, A191
Emsellem, E., Sutter, J., Chown, R., et al. 2026, arXiv:2607.03147
Engelbracht, C. W., Gordon, K. D., Rieke, G. H., et al. 2005, ApJL, 628, L29
Fitzpatrick, E. L., Massa, D., Gordon, K. D., Bohlin, R., & Clayton, G. C. 2019, ApJ, 886, 108
Galliano, F., Madden, S. C., Tielens, A. G. G. M., Peeters, E., & Jones, A. P. 2008, ApJ, 679, 310
Galliano, F., Galametz, M., & Jones, A. P. 2018, ARA&A, 56, 673
Genzel, R., Lutz, D., Sturm, E., et al. 1998, ApJ, 498, 579
Gleis, D. R., Stuber, S. K., Schinnerer, E., et al. 2026, A&A, 707, A121
Gordon, K. D., Bohlin, R., Sloan, G. C., et al. 2022, AJ, 163, 267
Gordon, K. D., Cartledge, S., & Clayton, G. C. 2009, ApJ, 705, 1320
Gordon, K. D., Clayton, G. C., Decleir, M., et al. 2023, ApJ, 950, 86
Gordon, K. D., Misselt, K. A., Bouwman, J., et al. 2021, ApJ, 916, 33
Gregg, B., Calzetti, D., Adamo, A., et al. 2026, ApJ, 997, 20
Groves, B., Krause, O., Sandstrom, K., et al. 2012, MNRAS, 426, 892
Groves, B., Kreckel, K., Santoro, F., et al. 2023, MNRAS, 520, 4902
Helou, G., Roussel, H., Appleton, P., et al. 2004, ApJS, 154, 253
Hunt, L. K., Thuan, T. X., Izotov, Y. I., & Sauvage, M. 2010, ApJ, 712, 164
Jarrett, T. H., Chester, T., Cutri, R., Schneider, S. E., & Huchra, J. P. 2003, AJ, 125, 525
Jarrett, T. H., Masci, F., Tsai, C. W., et al. 2013, AJ, 145, 6
Jensen, J. J., Hönig, S. F., Rakshit, S., et al. 2017, MNRAS, 470, 3071
Kaneda, H., Onaka, T., & Sakon, I. 2005, ApJL, 632, L83
Kennicutt, R. C., & Evans, N. J. 2012, ARA&A, 50, 531
Koornneef, J., Bohlin, R., Buser, R., Horne, K., & Turnshek, D. 1986, HiA, 7, 833
Lee, J. C., Sandstrom, K. M., Leroy, A. K., et al. 2023, ApJL, 944, L17
Leger, A., & Puget, J. L. 1984, A&A, 137, L5
Leroy, A. K., Sandstrom, K. M., Lang, D., et al. 2019, ApJS, 244, 24
Leroy, A. K., Sandstrom, K., Rosolowsky, E., et al. 2023, ApJL, 944, L9
Leroy, A. K., Schinnerer, E., Hughes, A., et al. 2021, ApJS, 257, 43
Li, A. 2020, NatAs, 4, 339
Maragkoudakis, A., Boersma, C., Peeters, E., et al. 2026, A&A, 709, A38
Maragkoudakis, A., Peeters, E., & Ricca, A. 2020, MNRAS, 494, 642
Mathis, J. S., Mezger, P. G., & Panagia, N. 1983, A&A, 128, 212
Meidt, S. E., Schinnerer, E., van de Ven, G., et al. 2014, ApJ, 788, 144
Muñoz-Mateos, J. C., Gil de Paz, A., Zamorano, J., et al. 2009, ApJ, 703, 1569
O'Halloran, B., Satyapal, S., & Dudik, R. P. 2006, ApJ, 641, 795
Panuzzo, P., Rampazzo, R., Bressan, A., et al. 2011, A&A, 528, A10
Pathak, D., Leroy, A. K., Barnes, A. T., et al. 2025, ApJL, 993, L20
Pathak, D., Leroy, A. K., Thompson, T. A., et al. 2024, AJ, 167, 39
Peeters, E., Hony, S., Van Kerckhoven, C., et al. 2002, A&A, 390, 1089
Peeters, E., Spoon, H. W. W., & Tielens, A. G. G. M. 2004, ApJ, 613, 986
Pessa, I., Schinnerer, E., Leroy, A. K., et al. 2022, A&A, 663, A61
Pessa, I., Schinnerer, E., Sanchez-Blazquez, P., et al. 2023, A&A, 673, A147
Querejeta, M., Meidt, S. E., Schinnerer, E., et al. 2015, ApJS, 219, 5
Querejeta, M., Schinnerer, E., Meidt, S., et al. 2021, A&A, 656, A133
Rampazzo, R., Panuzzo, P., Vega, O., et al. 2013, MNRAS, 432, 374
Rampazzo, R., Vega, O., Bressan, A., et al. 2014, A&A, 565, A50
Rigby, J., Perrin, M., McElwain, M., et al. 2023, PASP, 135, 048001
Rigopoulou, D., Spoon, H. W. W., Genzel, R., et al. 1999, AJ, 118, 2625
Salo, H., Laurikainen, E., Laine, J., et al. 2015, ApJS, 219, 4
Sandstrom, K. M., Bolatto, A. D., Draine, B. T., Bot, C., & Stanimirović, S. 2010, ApJ, 715, 701
Sandstrom, K. M., Koch, E. W., Leroy, A. K., et al. 2023, ApJL, 944, L8
Schinnerer, E., & Leroy, A. K. 2024, ARA&A, 62, 369
Skrutskie, M. F., Cutri, R. M., Stiening, R., et al. 2006, AJ, 131, 1163
Smith, J. D. T., Draine, B. T., Dale, D. A., et al. 2007, ApJ, 656, 770
Stierwalt, S., Armus, L., Surace, J. A., et al. 2013, ApJS, 206, 1
Sun, J., Teng, Y.-H., Chiang, I.-D., et al. 2025, ApJ, 994, 263
Sutter, J., Sandstrom, K., Chastenet, J., et al. 2024, ApJ, 971, 178

Tacconi-Garman, L. E., Sturm, E., Lehnert, M., et al. 2005, A&A, 432, 91
Tielens, A. G. G. M. 2005, The Physics and Chemistry of the Interstellar Medium (Cambridge Univ. Press)
Tielens, A. G. G. M. 2008, ARAA, 46, 289
van Diedenhoven, B., Peeters, E., Van Kerckhoven, C., et al. 2004, ApJ, 611, 928
Véron-Cetty, M.-P., & Véron, P. 2010, A&A, 518, A10
Whitcomb, C. M., Sandstrom, K., & Smith, J.-D. T. 2023, RNAAS, 7, 38
Whitcomb, C. M., Smith, J.-D. T., Sandstrom, K., et al. 2024, ApJ, 974, 20
Williams, T. G., Lee, J. C., Larson, K. L., et al. 2024, ApJS, 273, 13
Wolfire, M. G., Hollenbach, D., McKee, C. F., Tielens, A. G. G. M., & Bakes, E. L. O. 1995, ApJ, 443, 152
Wolfire, M. G., McKee, C. F., Hollenbach, D., & Tielens, A. G. G. M. 2003, ApJ, 587, 278
Xue, C., Byrne, A. N., Morgan, L., et al. 2025, ApJS, 281, 9
Zhang, L., & Ho, L. C. 2023, ApJ, 943, 60